\documentclass[aps,prd,floats,amsmath,amssymb,showpacs,nofootinbib]{revtex4}

\usepackage[utf8]{inputenc} 
\usepackage{latexsym} 
\usepackage{amsmath}  
\usepackage{amssymb}
\usepackage{hyperref}
\usepackage{color,graphicx}
\usepackage{epsfig}
\usepackage{booktabs}
\usepackage{graphicx}
\usepackage[caption=false]{subfig}

\begin{document}


\newcommand{\erik}[1]{\textcolor{blue}{{\bf Erik: #1}}}
\newcommand{\miguel}[1]{\textcolor{red}{{\bf Miguel: #1}}}


\title{Critical gravitational collapse of a massive complex scalar field}

\author{Erik Jim\'enez-V\'azquez}
\email{erjive@ciencias.unam.mx}

\author{Miguel Alcubierre}
\email{malcubi@nucleares.unam.mx}

\affiliation{Instituto de Ciencias Nucleares, Universidad Nacional
Aut\'onoma de M\'exico, A.P. 70-543, M\'exico D.F. 04510, M\'exico.}


\date{\today}


\begin{abstract}
We study the critical collapse of a massive complex scalar field coupled minimally to gravity. Taking as initial data a simple gaussian pulse with a shape similar to the harmonic ansatz for boson stars, we obtain critical collapse of type type I and II when varying the gaussian width $\sigma$. For $\sigma \leq 0.5$ we find collapse of type II with a critical exponent $\gamma=0.38\pm0.01$ and an echoing period $\Delta=3.4\pm0.1$. These values are very similar to the well known results for a real massless scalar field. On the other hand, for $\sigma \geq 2.5$ we obtain collapse of type I. In this case we find that the critical solutions turn out to be an unstable boson stars in the ground state: all the data obtained from our simulations can be contrasted with the characteristic values for unstable boson stars and their corresponding Lyapunov exponents.
\end{abstract}




\maketitle


\section{Introduction}

Due to the strong field dynamics, critical phenomena arises in gravitational collapse in the threshold of black hole formation. In a similar way to phase transitions in thermodynamics, taking the mass of black hole as an order parameter, critical gravitational collapse can be classified as type I or type II. In type I collapse the final black hole mass has a minimum finite value, whereas in type II collapse the black hole mass can be arbitrary small.
	
Historically, M. Choptuik discovered critical phenomena in gravitational collapse while studying numerically the gravitational collapse of a real massless scalar field, a result which was later named critical gravitational collapse of type II~\cite{PhysRevLett.70.9}. He found that, for a family of initial data parametrized by some arbitrary parameter $p$, the scalar field is completely dispersed to infinity for $p<p^*$, with $p^*$ some critical value, while for $p>p^*$ a black hole is formed with a final mass that follows a power-law scaling relation of the form:
\begin{equation}
M \propto (p-p^*)^\gamma \; .
\end{equation}
The critical solution $p=p^*$ that separates both states exhibits universality, {\em i.e.} it does not depend on the way in which the family is parameterized.  Additionally, for different types of matter the critical solution can have either continuous self-similarity (CSS), or discrete self-similarity (DSS). In particular, for the case of a real massless scalar field the critical solution was found to have DSS. This property is best appreciated in a logarithmic time defined as: 
\begin{equation}
T = -\ln ( \tau^*-\tau )  \; ,
\label{eq:log_time}
\end{equation}
with $\tau$ some measure of time which is usually taken as the proper time at the origin, and $\tau^*$ the so-called accumulation time. In this logarithmic time $T$ the solution is periodic with period $\Delta$. This property is known as ``scale echoing''. For the real massless scalar field the critical exponents have been found to be $\gamma \approx 0.374$ and $\Delta\approx3.445$, via both numerical simulations and semi-analytical studies \cite{PhysRevLett.70.9,Rinne:2020asi,PhysRevD.49.890,PhysRevD.51.5558,Hamad__1996,PhysRevD.92.084037}.
    
On the other hand, critical collapse of type I was later discovered by Choptuik et. al. while studying the critical collapse of a Yang-Mills field~\cite{PhysRevLett.77.424}. In contrast to the critical collapse of type II, in this case the final black hole mass has a minimum finite value, and there is a different scaling law of the form:
\begin{equation}
\label{eq:tau_scaling}
\tau \propto -\gamma \ln|p-p^*| \; ,
\end{equation}
where $\tau$ now measures the time that a given solution remains near the critical solution. Additionally, the critical solution itself is either stationary or periodic in time.
    
One can expect type I critical collapse when in the field equations there exists either a mass or a length scale that is relevant to the dynamics. On the contrary, when the equations do not contain a length scale, or when such a length scale is not relevant, type II phenomena occur. Both types can coexists in different regions in the parameter space of initial data, as for example in the case of a real massive scalar field~\cite{PhysRevD.56.R6057}, where the type of critical phenomena depends on the size of the Compton wavelength of the field when compared to the width of initial data (an excellent review about these type of critical phenomena can be found in~\cite{Gundlach:2007gc}).
   
In this paper we study the critical collapse of a massive complex scalar field. A previous study was done by Hawley and Choptuik in \cite{PhysRevD.62.104024}, showing that the critical solution for the complex scalar field corresponds to stationary solutions to the Einstein-Klein-Gordon known in the literature as boson stars~\cite{PhysRev.172.1331,PhysRev.187.1767,Visinelli:2021uve,Liebling:2012fv}. These solutions are determined by assuming spherical symmetry, and by the requirement that the metric coefficients must be static, while the complex scalar field has a harmonic time dependence of the form:
\begin{equation}\label{eq:BS_ansatz}
\Phi(t,r) = \varphi(r)e^{i\omega t} \; ,
\end{equation}
with $\omega$ a real valued frequency, and $\varphi(r)$ a real valued function of radius only. Taking the mass parameter of the complex scalar field as $m$, the maximum possible mass of a boson star has been found to be $M_{max}\approx0.633 \: M_{Planck}^2/m$, corresponding to a central value of the scalar field of $\varphi_{max} = \varphi(0) \approx 0.271$, see for example~\cite{PhysRevD.42.384,GLEISER1989733} (though this value can change depending on the normalization, see below for our normalization choice).  This central value of the field separates the boson star configurations into two branches depending on their stability properties. If $\varphi(0)<\varphi_{max}$ the boson star is stable under small perturbations, whereas for $\varphi(0)>\varphi_{max}$ the configurations are unstable. The lowest energy solution for a boson star for a given value of $\varphi(0)$, also known as the ground state, has no nodes in the scalar field. Excited states are classified depending on the number nodes of the field in the radial direction.

The critical solutions found by Hawley and Choptuik corresponded to unstable boson stars in the ground state, and were obtained by perturbing a stable boson star that interacted gravitationally with a small pulse of a massless scalar field that acts as the perturbation. In our study we take a different approach, and we begin with a simple gaussian pulse in the complex field with a variable width. We then evolve this initial data and vary the amplitude of the gaussian pulse until a critical solution is found. 
    
Since the mass of the complex scalar field introduces a scale, we expect our system to display both types of critical phenomena depending on the width of the initial gaussian pulse. In a similar way as in the case of a real massive scalar field~\cite{PhysRevD.56.R6057}, we will explore both types of critical behaviour by changing the width of our initial pulse. Furthermore, if one performs a linear perturbative analysis for critical phenomena of both type I and II, the critical exponent $\gamma$ can be shown to be the inverse of the so-called Lyapunov exponent $\chi$ of the system $\gamma=1/\chi$~\cite{Gundlach:2007gc}.\footnote{The Lyapunov exponent measures the stability of a system due to changes in its initial conditions. For close trajectories in phase space parametrized by $t$, and initial points separated by an infinitesimal distance $\delta$, the Lyapunov exponent quantifies their rate of separation as $F(t,x_0+\delta)-F(t,x_0)\approx\delta e^{\chi t}$, with $\chi$ the Lyapunov exponent of the system. For $\chi>0$ the trajectories diverge, whereas for $\chi<0$ they do not.}

For the case of critical collapse of type I, we will compare our critical solutions with the known solutions for stationary boson stars. Furthermore, we can also compare our critical exponents with the Lyapunov exponents for the unstable modes of boson stars. On the other hand, when the critical phenomena is of type II we limit our study to finding the critical exponent $\gamma$ and the echoing period $\Delta$.
    
This paper is organized as follows.  In Section~\ref{sec:scalarfield} we discuss the field equations for a complex massive scalar field as weel as our initial data. In Section~\ref{sec:code} we discuss our numerical code, gauge conditions and diagnostic tools. Section~\ref{sec:results} shows the results of our numerical simulations.


\section{Complex scalar field}
\label{sec:scalarfield}


\subsection{The Einstein Klein-Gordon equations}

Our matter model consists of a massive complex scalar field $\Phi$ coupled minimally to gravity, which can be described by the action (in units such that $G=c=1$):
\begin{equation}
S = \int d^4x \sqrt{-g} \left[ \frac{R}{16\pi}
- \frac{1}{2} \left( \nabla^\mu\Phi\nabla_\mu \Phi^*
+ m^2\Phi\Phi^* \right) \right] \; ,
\end{equation}
where $R$ is the Ricci scalar of the spacetime and $m$ is the mass parameter of the complex scalar field (notice that this fixes our normalization choice). Varying the action with respect to the metric and the scalar field one obtains the Einstein field equations:
\begin{equation}
R_{\mu \nu} - \frac{1}{2} g_{\mu \nu} R = 8 \pi T_{\mu \nu} \; ,
\end{equation}
together with the Klein--Gordon equation:
\begin{equation}\label{eq:KG}
\nabla^{\mu} \nabla_{\mu} \Phi-m^{2} \Phi=0 \; ,
\end{equation}
where the stress-energy tensor $T_{\mu \nu}$ for the scalar field is given by:
\begin{equation}
T_{\mu\nu} = \frac{1}{2} \left[ \left( \nabla_{\mu} \Phi \nabla_{\nu} \Phi^{*}
+ \nabla_{\nu} \Phi \nabla_{\mu} \Phi^{*} \right)
- g_{\mu\nu} \left( \nabla^\alpha \Phi \nabla_\alpha \Phi^* + m^2 \Phi \Phi^* \right) \right] \; .
\end{equation}

In order to study numerically the evolution of the system, we will use the Baumgarte-Shapiro-Shibata-Nakamura (BSSN) formulation of general relativity \cite{PhysRevD.52.5428,PhysRevD.59.024007}, which is known to be strongly hyperbolic \cite{PhysRevD.66.064002}. Particularly, as we are only interested in the case of spherical symmetry, we will use the BSSN formulation adapted to curvilinear coordinates as described in~\cite{PhysRevD.79.104029,Alcubierre:2011pkc}. In spherical symmetry, we will adopt the line element given by:
\begin{equation}
\label{eq:ds4}
ds^2 = -\alpha^2 dt^2 + \psi^4 \left( A dr^2 + r^2 B d \Omega^2 \right) \; ,
\end{equation}
where $(\alpha,\psi,A,B)$ are functions of $(t,r)$ only, and $d\Omega^2=d\theta^2+\sin^2\theta\ d\varphi^2$ is the standard solid angle element.

In order to recast the Klein--Gordon equation as a first order system we define the following auxiliary variables
\begin{equation}
\Pi := \frac{\partial_t\Phi}{\alpha} \; , \qquad \chi := \partial_r \Phi \; .
\end{equation}
With these definitions, the Klein--Gordon equation \eqref{eq:KG} can be rewritten as:
\begin{eqnarray}
\partial_t\Phi &=& \alpha\Pi \; , \\
\partial_t\chi &=& \alpha \partial_r \Pi + \Pi \partial_r \alpha \; , \\
\partial_t\Pi &=& \frac{\alpha}{A\psi^4} \left[ \partial_{r} \chi
+ \chi \left( \frac{2}{r} - \frac{\partial_{r} A}{2 A} + \frac{\partial_{r} B}{B}
+ 2 \partial_{r} \ln \psi \right) \right]
+ \frac{\chi \partial_{r} \alpha}{A \psi^{4}} + \alpha K \Pi - \alpha m^2 \phi \; ,
\end{eqnarray}
with $K:=K^m_m$ the trace of the extrinsic curvature of the spatial hypersurfaces of constant time.

From the orthogonal decomposition of the stress-energy tensor:
\begin{equation}
T^{\mu \nu} = S^{\mu \nu} + J^{\mu} n^{\nu} + n^{\mu} J^{\nu} + \rho n^{\mu} n^{\nu} \;,
\end{equation}
we obtain the energy density $\rho:=n^{\mu} n^{\nu} T_{\mu \nu}$, the momentum density $J_\mu:= -P_{\ \mu}^{\nu} n^{\lambda} T_{\nu \lambda}$, and the stress tensor \mbox{$S_{\mu \nu}:=P_{\mu}^{\sigma} P_{\nu}^{\lambda} T_{\sigma \lambda}$}, where $n^\mu=(1/\alpha,0,0,0)$ is the unit normal vector to the spatial hypersurfaces and $P^\mu_{\ \nu}:=\delta^\mu_{\ \nu}+n^\mu n_\nu$ is the projection operator.  For the complex scalar field we find in particular:
\begin{eqnarray}
\rho &=& \frac{1}{2} \left( |\Pi|^{2} + \frac{|\chi|^{2}}{A \psi^{4}} + m^2|\Phi|^2 \right) \; , \\
J_r &=& -\frac{1}{2} \left( \rule{0mm}{5mm} \chi \Pi^{*} + \Pi \chi^{*} \right) \; , \\
S_r^r &=& \frac{1}{2} \left( |\Pi|^{2} + \frac{|\chi|^{2}}{A \psi^{4}} - m^2|\Phi|^2 \right) \; , \\
S_{\theta}^{\theta} &=& \frac{1}{2} \left( |\Pi|^{2} - \frac{|\chi|^{2}}{A \psi^{4}}
- m^{2}|\Phi|^{2} \right) \; .
\end{eqnarray}


\subsection{Initial data}

In \cite{PhysRevD.62.104024} Hawley and Choptuik showed that the critical solution for the case of a massive complex scalar field is an unstable boson star. In their study the critical solution was obtained by perturbing a boson star in the stable branch with a real massless scalar field, and tuning the amplitude of the massless field up to the threshold of black hole formation. Here we will take a different approach, by considering as our initial condition a simple pulse of complex scalar field with the following gaussian profile:
\begin{eqnarray}
\Phi(t=0,r) &=& \Phi_0 e^{-r^2/\sigma^2} \label{eq:phi} \; , \\
\Pi(t=0,r) &=& i \kappa \Phi_0 e^{-r^2/\sigma^2} \label{eq:pi} \; ,
\end{eqnarray}
where $\Phi_0,\sigma,\kappa$ are real parameters, and $\Phi_0$ is the tuning amplitude of the initial pulse. In order to find the initial data for the geometry we assume a conformally flat spatial metric, so that $A=B=1$ in~\eqref{eq:ds4}, and proceed to solve the constraint equations. Notice that even though equations \eqref{eq:phi} and \eqref{eq:pi} do not formally represent an instant of time symmetry, the momentum density $J_r$ is still zero, so the momentum constraint is trivially satisfied. On the other hand, at $t=0$ the hamiltonian constraint becomes a nonlinear second order differential equation for the conformal factor $\psi$ of the form:
\begin{equation}
\partial^2_r\psi + \frac{2}{r} \: \partial_r \psi + 2 \pi \psi^5 \rho = 0 \; ,
\label{eq:ham_const}
\end{equation}
with the energy density given by:
\begin{equation}
\rho = \frac{1}{2} \left[ | \Pi |^2 + \frac{|\partial_r \Phi|^2}{\psi^4} + m^2 |\Phi|^2 \right] \; .  
\end{equation}
The above non-linear equation is solved numerically by using an iterative method. Boundary conditions for equation \eqref{eq:ham_const} are obtained from the asymptotically flatness condition, which implies:
\begin{equation}
\psi(r)|_{r\rightarrow\infty} = 1 \; .
\end{equation} 
In practice, however, we use a Robin boundary type condition at a finite radius corresponding to the edge of our numerical grid:
\begin{equation}
\partial_r\psi = \frac{1-\psi}{r} \; .
\end{equation}
This condition reflects the fact that as $r\rightarrow\infty$ we have $\psi\rightarrow1+{\cal O}(r^{-1})$. On the other hand, regularity at the origin implies that $\psi$ must be an even function of $r$, so that:
\begin{equation}
\partial_r \psi|_{r=0} = 0 \; .
\end{equation}
The initial conditions given by equations~\eqref{eq:phi}-\eqref{eq:pi} were not chosen randomly. At $t=0$ they are similar to the harmonic ansatz for boson stars, Eq.~\eqref{eq:BS_ansatz}, with $\kappa$ taking the role of the oscillation frequency $\omega$ of the boson star, but instead of using the profile for a stable boson star we use a simple gaussian profile. Notice that $\kappa$ is a free parameter, which we later choose to be equal to the mass $m$ of the scalar field $\kappa=m$ for simplicity.


\section{Numerical code and diagnostics}
\label{sec:code}

\subsection{Code and gauge choices}

We integrate the Einstein--Klein--Gordon system with the OllinSphere code, a numerical relativity finite-difference code suited for spherical symmetry which evolves the BSSN formulation of the Einstein equations. This code has been previously used for example in~\cite{Alcubierre:2019qnh,Degollado:2020lsa}, but it has now been updated to include the possibility of fixed mesh refinement around the origin, so that if the outer boundary is located at $r_{max}$ with grid resolution $\Delta r$, the local boundary of the $N$ refinement level is situated at $r_{max}/2^{N-1}$ with  resolution $\Delta r/2^{N-1}$, and the grid structure remains fixed during the evolution.

To close the system we also need to specify the lapse function $\alpha$.  In our simulations we choose for the lapse the standard $1+\log$ slicing condition:
\begin{equation}
\partial_t\alpha = -2 \alpha K \; ,
\end{equation}
with $K$ the trace of the extrinsic curvature. We choose for the initial value of the lapse a pre-collapsed profile of the form $\alpha(t=0)=\psi^{-2}$, with $\psi$ the initial conformal factor. Also, for simplicity we choose a vanishing shift vector for all our simulations. Indeed, this choice was already assumed in the line element~\eqref{eq:ds4}.


\subsection{Diagnotics}

As is usual in the study of critical behavior, the final state of the evolution is classified depending on the strength of the initial data. We report our initial amplitude precision in finding the critical solution via the dimensionless quantity: 
\begin{equation}
\delta\Phi = \frac{\Phi_c - \Phi_d}{\Phi_d} \; ,
\end{equation}
where $\Phi_c$ is the highest amplitude for which the initial data is dispersed and leaves behind Minkowski spacetime, and $\Phi_d$ is the lowest amplitude for which a black hole is formed. The critical value for the amplitude $\Phi^*$ is a metastable state which separates the two behaviors described previously. We have found that in order to obtain the critical exponents correctly we need an accuracy of at least $\delta\Phi\sim10^{-6}$, o even higher. This value can be improved by using a finer grid, and results in less uncertainty in the value of the critical exponents, and also in a longer evolutions near the critical solution for type I critical collapse.

Since we are mostly interested in the subcritical case, for the supercritical simulations we will not follow the evolution until the black hole settles down to equilibrium, which in any case would require a non-zero shift vector. With our slicing choice, the lapse function at the origin will return to one if the initial scalar field pulse is dispersed to infinity. Otherwise, if a black hole is formed, the lapse will collapse to zero at the origin.

In order to detect when a black hole is formed we will search at every time step for an apparent horizon. This procedure is done by looking for a location where the expansion of outgoing null geodesics becomes zero (see for example~\cite{Alcubierre:1138167}):
\begin{equation}
\frac{1}{\psi^{2} \sqrt{A}} \left( \frac{2}{r} + \frac{\partial_{r} B}{B}
+ 4 \frac{\partial \psi}{\psi} \right) - 2 K_{\theta}^{\theta} = 0 \; .
\end{equation}
Here $K_\theta^\theta$ is simply the angular component of the extrinsic curvature with mixed indices.


\subsection{Characterizing type I critical solutions}

In Eq.~\eqref{eq:tau_scaling} we take $\tau$ as the proper time measured by an observer located at the origin $r=0$ at a point in the evolution when a first apparent horizon is located. As explained before, the critical solutions of type I for the complex scalar field should correspond to an unstable boson star. We should emphasize, however, that our initial conditions for the complex scalar field given by Eqs.~\eqref{eq:phi}-\eqref{eq:pi} with the critical amplitude $\Phi_0^*$ do not correspond to a boson star at $t=0$, unstable or otherwise. This implies that for our near critical simulations the excess of scalar field will be radiated to infinity and the remaining content should approach a boson star in the unstable branch.

Since boson stars do not have a well defined boundary, one can describe their size by means of the so-called $R_{95}$ or $R_{99}$ radius, which correspond to the areal radius of a sphere containing $95\%$ or $99\%$ of the total mass $M_T$, respectively. Furthermore, since the system is not stationary the integrated mass will be a function of $M(t,r)$. To determine if a compact object has formed, we inspect the compactness function defined as:
\begin{equation}\label{eq:compactness}
C(t,r) = \frac{M(t,r)}{R(t,r)} \; ,
\end{equation}
where $R(t,r)$ is the areal radius of a sphere at a given time $t$ and coordinate radius $r$. We look for the global maximum as a function of $r$ for every time step. We expect that if a boson star has formed, there will be a maximum mean value of $C(t,r)$, plus some small oscillations around it corresponding to perturbations of this star. This behavior will tell us if a compact object has formed or not, and will also provide us with an approximate lifetime of the critical solution obtained.

In Eq.~\eqref{eq:compactness} we estimate the mass function $M(r,t)$ of the configuration by using the Kodama mass~\cite{Kodama:1979vn, PhysRevD.91.084057,PhysRevD.96.064047,Racz:2005pm}, which is a quasi-local conserved energy in a spherically symmetric spacetime. The Kodama vector is defined by:
\begin{equation}
K^A = \epsilon^{AB} \partial_{B} R \; ,
\end{equation}
where $R$ is the areal radius of a sphere at constant $t$ and $r$, $\epsilon^{A B}$ is the totally antisymetric tensor in the two-dimensional manifold with coordinates $(t,r)$, and the indices $(A,B)$ run over $(0,1)$. The vector $K^A$ can be naturally extended to the four-dimensional manifold  
by setting to zero the remaining components. Next, we define the four vector $S^\mu$ as follows:
\begin{equation}
{\cal{S}}^\mu = T^{\mu \nu} K_\nu \; ,
\end{equation}
where $T^{\mu \nu}$ is the stress-energy tensor. It is possible to show that ${\cal S}^\mu$ is a conserved current, so it satisfies the conservation law:
\begin{equation}
\partial_\mu \left( \sqrt{-g} {\cal S}^\mu \right) = 0 \; ,
\end{equation}
In a sphere of radius $r$ at constant $t$, we can then define a conserved mass, the so-called Kodama (or Misner--Sharp)q mass as:
\begin{equation}
M\left(t, r\right) := \int_{\text {sphere }} {\cal S}^{t} \alpha \sqrt{\gamma} \:d x^{3} \;,
\end{equation}
where $\gamma$ is the determinant of the 3-metric, and where we used the fact that $-g = \alpha \gamma$. Using our expression for the spatial metric this reduces to:
\begin{equation}\label{eq:M_Kodama}
M(t,r) := 4 \pi \int_0^r  \alpha {\cal S}^{t} r^2 \psi^6 A^{1/2} B \: dr \; ,
\end{equation}
Notice that the above expression allows us to have local concept of mass as a function of $r$ and $t$.\footnote{In spherical symmetry the are other equivalent forms of calculating a local mass.  For example, one can write the radial metric in terms of the areal radius $r_a$ as $g_{rr}=1/(1-2m(r_a)/r)$ and solve for $m(r_a)$.  We prefer the integral above as it depends directly on the stress-energy tensor and in practice seems to be less prone to numerical errors.}


In order to find the total mass of the unstable boson star corresponding to the critical solution we still need to estimate its radius $R$. Notice that we can not simply calculate the integral~\eqref{eq:M_Kodama} all the way to the boundary boundary of the numerical grid, since some of the initial scalar field will be continuously radiated away and should not be considered as part of the critical solution.  To estimate the radius $R$ we use the fact that stationary boson stars are well characterized in the literature, and our code is capable of finding those solutions (see for example~\cite{Alcubierre:2018ahf}).  For a near critical simulation we then first obtain the mean value of the scalar field amplitude at the origin, $\left< \Phi(t,r=0) \right>$. Having found this mean amplitude, we construct the corresponding stationary boson star solution with that same amplitude, and choose $R$ as the $R_{99}$ radius of that stationary solution.


\subsection{Characterizing type II critical collapse}

As the mass of the scalar field introduces a length scale, following \cite{PhysRevD.56.R6057} we expect that if $\sigma m \ll 1$ we should observe critical collapse of type II, whereas if $\sigma m \gg 1$ we should find collapse of type I.  Finding the value of the critical exponent $\gamma$ using the final black hole mass scaling is somewhat difficult since we would need to follow the black hole until it reaches an equilibrum configuration, something that is not trivial to do numerically. Instead, we will consider subcritical evolutions since in a critical collapse of type II the maximum value of the 4D Ricci scalar will then follow the scaling law:
\begin{equation}
R_{max} \approx |\Phi_0^*-\Phi_0|^{-2\gamma} \; ,
\end{equation}
where the $-2$ factor in the exponent is there because the Ricci scalar has units of lenght$^{-2}$. Additionally to this behavior, the discrete self-similarity of the phenomena adds a fine structure to the scaling law \cite{PhysRevD.55.R440}, so the Ricci scalar in fact will behave as:
\begin{equation}
\label{eq:4DR_1}
\ln R_{max} = c - 2 \gamma \ln |\Phi_0^* - \Phi_0| + f( \ln |\Phi_0^*-\Phi_0| ) \; ,
\end{equation}
with $c$ some constant, and where $f$ is a periodic function with angular frequency:
\begin{equation}
\omega = \Delta /2 \gamma \; ,
\label{eq:Delta_fit}
\end{equation}
where $\Delta$ is the so-called echoing period. To leading order, $f$ can be approximated by:
\begin{equation}
f(x) = a_0 \sin \left( \omega x + \varphi \right) \; ,
\end{equation}
with $\varphi$ some arbitrary phase. The 4D Ricci scalar then behaves as:
\begin{equation}
\label{eq:4DR_2}
\ln R_{max} = c - 2\gamma \ln |\Phi_0^*-\Phi_0|
+ a_0 \sin \left( \omega \ln |\Phi_0^*-\Phi_0| + \varphi \right) \; ,
\end{equation}
where the constants $c,a_0,\varphi$ depend on the form of the initial data family.

A second method to obtain $\Delta$ was described in \cite{PhysRevD.98.084012}. Originally, this method was applied to the case of a real massless scalar field, and uses the fact that critical solution is periodic in the logarithmic time $T$. Here we will apply this method to the case of a complex scalar field by considering the proper time for two pairs of consecutive local minima of the magnitude of the scalar field $\|\Phi\|$ evaluated at the origin, $(\tau_n,\tau_{n+1})$ and $(\tau_m,\tau_{m+1})$, which corresponds to the pairs $(T_n,T_{n+1})$, $(T_m,T_{m+1})$ in the logarithmic time.  Assuming now that each pair differs in half of the period $\Delta/4$, one can solve for the accumulation time $\tau^*$ obtaining:
\begin{equation}\label{eq:acc_t}
\tau^{*}=\frac{\tau_{n} \tau_{m+1}-\tau_{n+1} \tau_{m}}{\tau_{n}-\tau_{n+1}-\tau_{m}+\tau_{m+1}} \, .
\end{equation}
This procedure also provides us with an estimate of the echoing period $\Delta$ given by:
\begin{equation}\label{eq:delta_echo}
\Delta = 2 \ln \left(\frac{\tau^{*}-\tau_{n}}{\tau^{*}-\tau_{n+1}}\right) \,.
\end{equation}


\section{Numerical results}
\label{sec:results}

All our simulations were performed with fourth order centered differences in space, and fourth order Runge--Kutta for the evolution in time. For simplicity, we fix the scalar field mass to $m=1$. Also, for the family of initial data~\eqref{eq:phi}-\eqref{eq:pi} we set $\kappa=m=1$ for all cases. The values chosen for the width parameter $\sigma$ will be reported below. The parameter $\Phi_0$ is then adjusted until we find the black hole formation threshold with the desired accuracy.

To reduce the source of errors in our simulations we use constraint preserving boundary conditions. These have already been described and used for example in~\cite{Alcubierre:2014joa,PhysRevD.86.104044}, and they help to reduce the errors coming in from the boundaries by a factor of about $10^3$ when compared with the standard Sommerfeld (radiative) boundary conditions. The error introduced by the finite difference method can also be diminished by using Kreiss--Oliger numerical dissipation. In all our evolutions we use sixth-order dissipation in order to be compatible with the fourth-order discretization. The artificial dissipation dampens high frequency modes that would otherwise spoil the numerical stability of the near-critical solutions. Resolution also affects the critical behavior (in particular the precise value of the critical amplitude $\Phi_0^*$), for this reason we will report relevant quantities for our highest resolution simulations.


\subsection{Type II critical collapse}

As already stated before, we expect type II critical collapse for $\sigma m \ll 1$. To check this, we choose $\sigma\leq0.5$ and proceed to find the critical amplitude $\Phi_0^*$ using a bisection method. Since studying critical phenomena requires high numerical precision, instead of using many levels of refinement, which introduce reflections at the refinement boundaries, we will use just one grid level with a radial transformation of coordinates from the original coordinate radius $r$ to a new coordinate $\tilde{r}$ related to $r$ through:
\begin{equation}
\label{eq:rnew}
\frac{dr}{d\tilde{r}} = \frac{1}{1+e^{\beta r^2+\delta}} \; .
\end{equation}
With this transformation a uniform grid in $\tilde{r}$ becomes a non-uniform grid in $r$. This coordinate transformation was first used in~\cite{PhysRevD.105.064071} for studying the critical behavior of scalar-tensor theories of gravity in the Jordan frame. In Eq. \eqref{eq:rnew}, $\delta$ adjusts the resolution near the origin $\tilde{r}=0$, while $\beta$ measures how fast $\tilde{r}$ approaches $r$ far away. Notice that as the transformed radial coordinate approaches infinity $\tilde{r}\rightarrow\infty$, we have $dr/d\tilde{r}\rightarrow 1$. For our simulations we use $\delta=5$ and $\beta=-1$, with a grid spacing $\Delta \tilde{r}=0.005$, and $N_r=2500$ points in radial direction. We also use an adaptive time step in order to satisfy the Courant--Friedrichs--Levy (CFL) stability condition. With these settings we were able to find the critical amplitude with a precision of $\delta\Phi\approx10^{-12}$.

\begin{figure}
\includegraphics[width=0.75\textwidth]{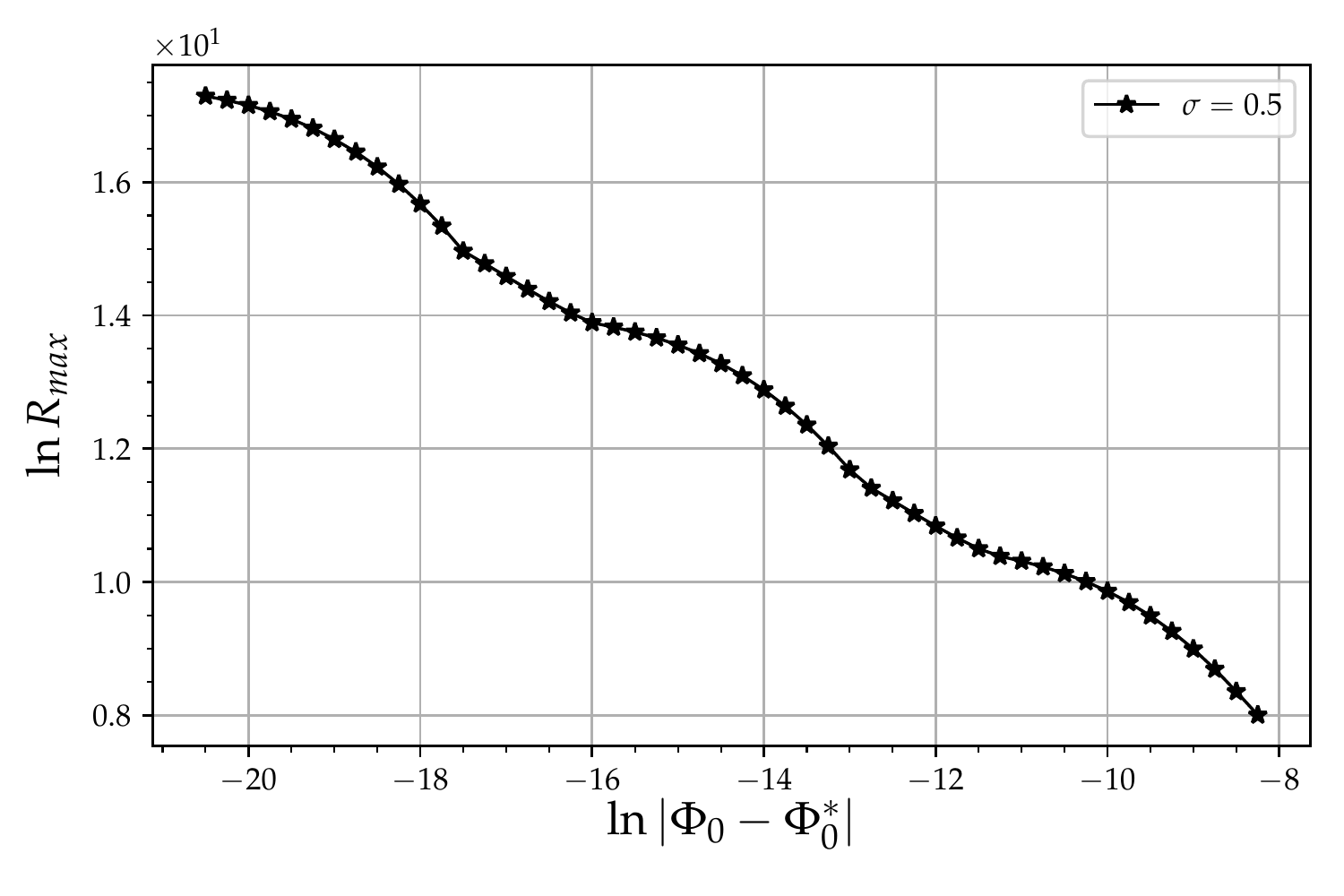}
\caption{Scaling of the maximum value of the 4D Ricci scalar for subcritical simulations of a massive complex scalar field, using the initial data family \eqref{eq:phi}-\eqref{eq:pi}, with gaussian width $\sigma=0.5$.}
\label{fig:lnRmax}
\end{figure}

Figure~\ref{fig:lnRmax} shows the maximum value of the 4D Ricci scalar at the origin obtained from subcritical evolutions for the particular case $\sigma=0.5$. From the Figure we can clearly see the expected behaviour for type II critical collapse. Table \ref{tab:all-exponents} shows the critical exponents obtained for the different values of $\sigma=0.2,0.3,0.4,0.5$. As we can see, for all these cases fitting the function \eqref{eq:4DR_2} results in critical exponents that are very close to those found in the literature for the case of a real massless scalar field.

\begin{figure}
\includegraphics[width=0.75\textwidth]{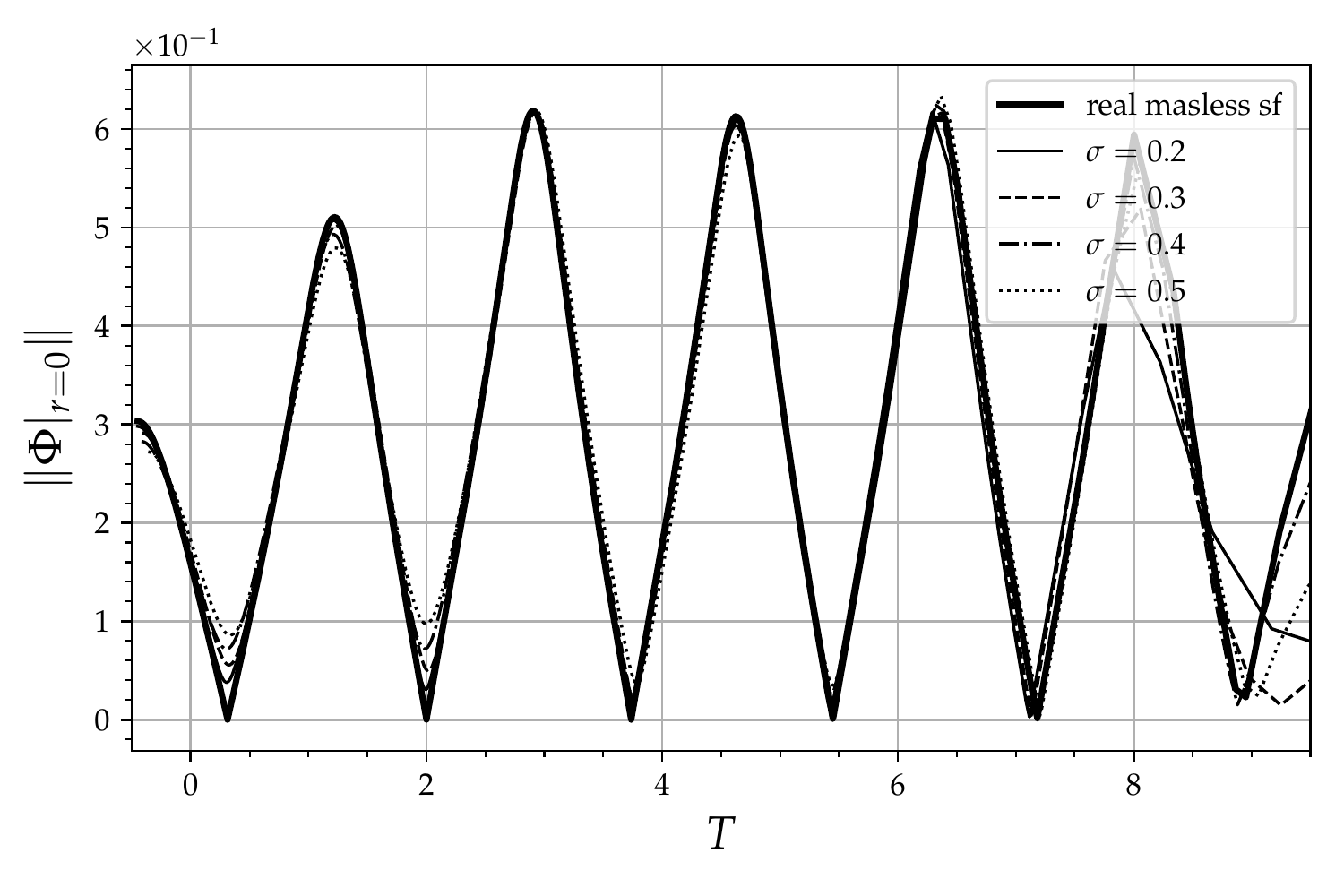}
\caption{Central value of the norm of the complex massive scalar field compared with the real massless case. Plots have been shifted in time in order to coincide with the real case. As expected, since $\Delta$ are very similar to each other, all lines overlap with the real case.}
\label{fig:norm_phi}
\end{figure}

In Figure \ref{fig:norm_phi}, for near-critical evolutions we compare the magnitude of central value of the massive complex scalar field, with that of a real massless scalar field in logarithmic time $T$. The overlap we find is no surprise since the values of $\gamma$ and $\Delta$ in all cases are very similar to the critical exponents for the Choptuik solution.
\begin{table}
\centering
\begin{tabular}{|c|c|c|c|}
\hline
$\sigma$ & $\gamma$ & $\Delta \eqref{eq:Delta_fit}$ & $\Delta \eqref{eq:delta_echo}$       \\ \hline
0.2      & 0.374$\pm$0.001 & 3.423 $\pm$ 0.026 & 3.426 $\pm$ 0.026 \\
0.3      & 0.375$\pm$0.001 & 3.442 $\pm$ 0.025 & 3.424 $\pm$ 0.031 \\
0.4      & 0.376$\pm$0.001 & 3.493 $\pm$ 0.019 & 3.442 $\pm$ 0.033 \\
0.5      & 0.376$\pm$0.001 & 3.440 $\pm$ 0.021 & 3.436 $\pm$ 0.059 \\ \hline
\end{tabular}
\caption{Summary of all exponents obtained for cases $\sigma\leq0.5$. Up our uncertainties, they are very similar to the values found in the literature for the case of a real massless scalar field.}
\label{tab:all-exponents}
\end{table}

\subsection{Type I critical collapse}

As already mentioned, for the case when $\sigma m\gg 1$ we expect to find type I critical collapse. To investigate this, we start from $\sigma=2.5$ and consider higher values of the gaussian width. Once the value of $\sigma$ has been chosen, for each case we proceed to find the critical amplitude $\Phi_0^*$ with the bisection method. For these simulations we have used fixed mesh refinement with $N=4$ levels. We have observed that the lifetime of the nearest critical solution obtained initially increases as $\sigma$ is increased, reaching its maximum for $\sigma \approx 4.0$, and then decreases again for higher values of $\sigma$.

To fix the position of the outer boundary, we first estimate the time of black hole formation $t_{BH}$ using low resolution runs. We then multiply $t_{BH}$ the asymptotic gauge speed $v_g=\sqrt{2}$ of the $1+\log$ slicing condition (as it is larger than the coordinate speed of light $v_l=1$). A perturbation that starts at the origin and bounces at the boundary will take at least twice this time to return to the origin (in fact longer since the lapse is smaller than one near the origin), so we set the outer boundary at $r_{max}=t_{BH}/\sqrt{2}$. The position of the outer boundary $r_{max}$, and the resolution of the base grid $\Delta r$, are displayed in Table~\ref{tab:resolution_bdry}. In all these simulations we use a fixed time step compatible with the CFL condition.

\begin{table}
\begin{tabular}{@{}|c|c|c|@{}}
	\hline
	\multicolumn{1}{|c|}{$\sigma$} & \multicolumn{1}{c|}{$\Delta r$} & \multicolumn{1}{c|}{$r_{max}$ } \\ \hline
	2.5      &            & 150       \\
	2.75     & 0.1        & 200       \\
	3        &            & 250       \\ \hline
	3.5      &            & 400       \\
	4        & 0.15       & 550      \\
	5        &            & 400       \\ \hline
	6        &            & 350       \\
	7        &            & 325       \\
	8        & 0.1        & 325       \\
	9        &            & 290       \\
	10       &            & 225       \\ \hline
\end{tabular}
\caption{Resolution and position of the outer boundary for each value of the width parameter $\sigma$ for our simulations of type I critical collapse.}
\label{tab:resolution_bdry}
\end{table}

\begin{figure}
\includegraphics[width=0.75\textwidth]{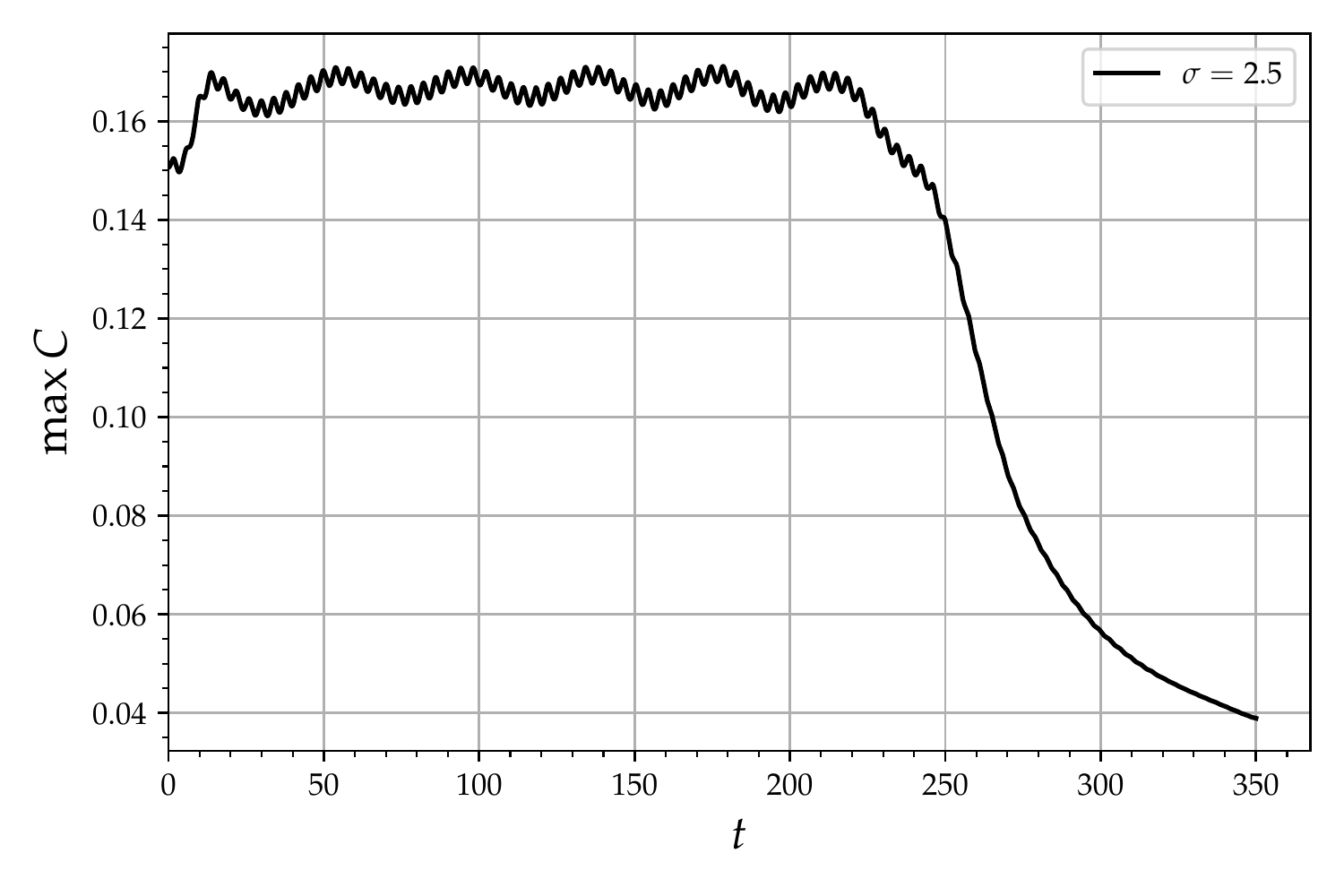}
\includegraphics[width=0.75\textwidth]{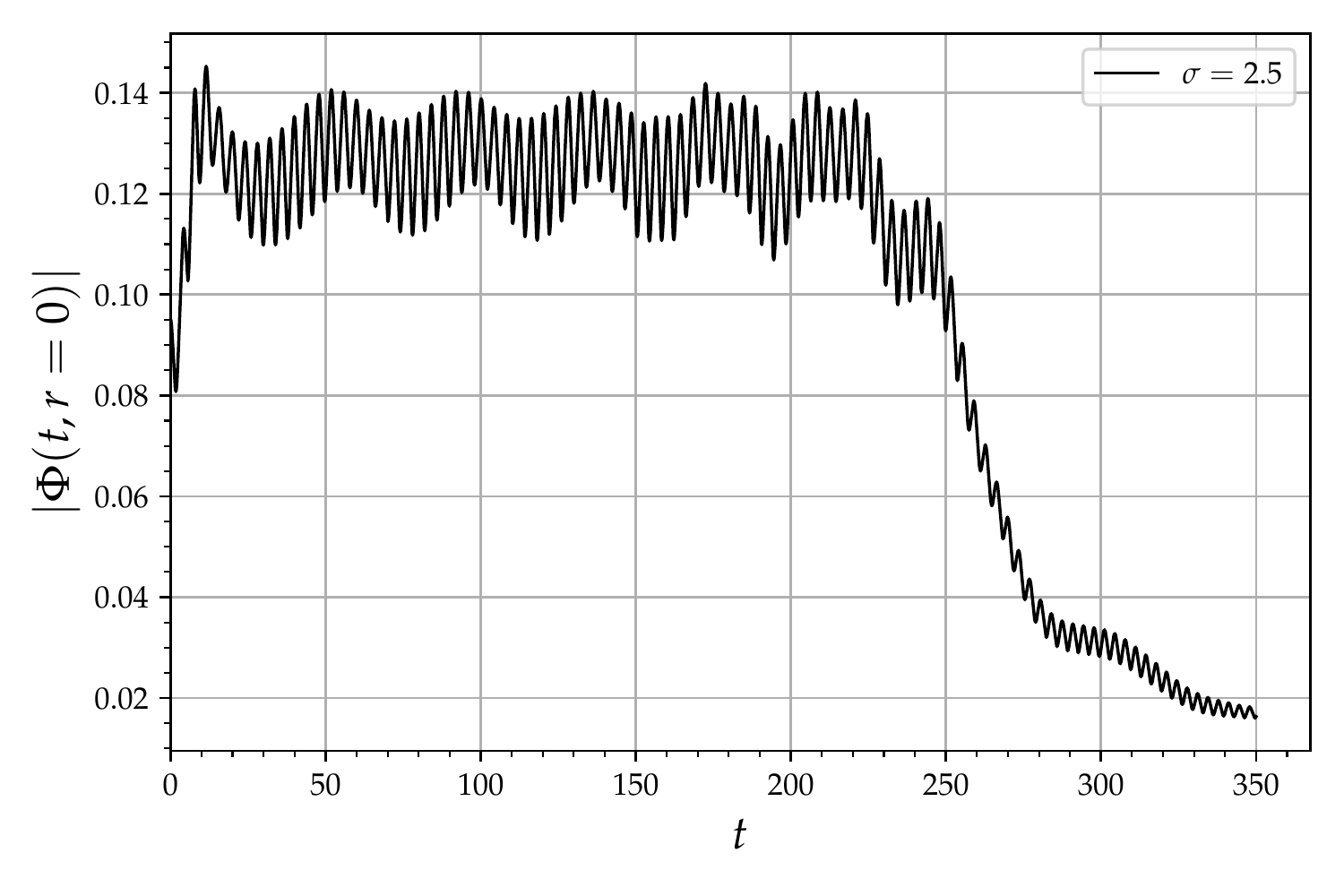}
\caption{Top panel: Maximum value of the compactness function for a near critical evolution with $\sigma=2.5$. Bottom panel: Norm of the scalar field at the origin.  At very early times we see no indication that a compact object has formed. However, from $t \approx 25$ up to $t \approx 225$ we can appreciate a clear oscillation around a mean value. Since this is a subcritical case, we see dispersion of the object for $t > 225$.}
\label{fig:compactness-norm}
\end{figure}

Figure~\ref{fig:compactness-norm} shows the maximum value of the compactness function (top panel), and the norm of the complex scalar field at the origin (bottom panel), for a near critical solution with initial width $\sigma=2.5$. Since the initial profile consists of a gaussian pulse, the scalar field has not agglomerated to form a compact object. After $t \approx 25$ a portion of the scalar field has been radiated away and the remainder starts to oscillate around a mean value, indicating that a compact object has been formed. As this state is unstable, the object eventually disperses after $t \approx 225$. It is important to mention, however, that for some values of the initial amplitude $\Phi_0$ the scalar field does not disperse completely, and the remaining bulk oscillates around the origin. This behavior is similar to that observed by Lai and Choptuik in~\cite{Lai:2007tj}. In their study, the remaining bulk can be described as excitations of the fundamental mode of stable boson stars. However we will leave the study of this phenomenon for a future work.

\begin{figure}
\includegraphics[width=0.75\textwidth]{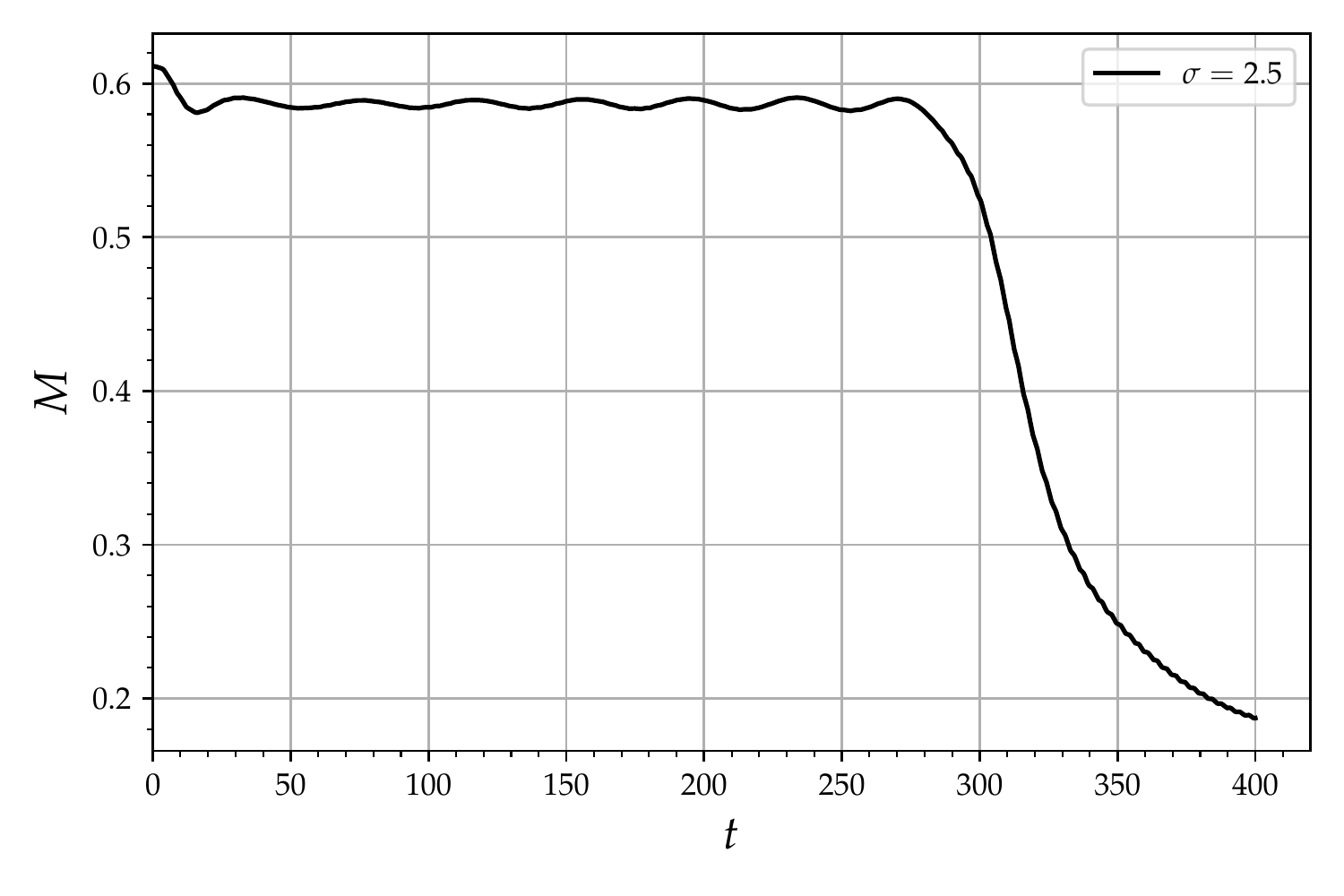}
\caption{Kodama mass for the same simulation of Figure~\ref{fig:compactness-norm}. After obtaining the mean value of the norm of the scalar field at the origin we find the $R_{99}$ of the corresponding boson star. We then evaluate the Kodama mass at that radius.}
\label{fig:KodamaMass}
\end{figure}

\begin{figure}
\includegraphics[width=0.75\textwidth]{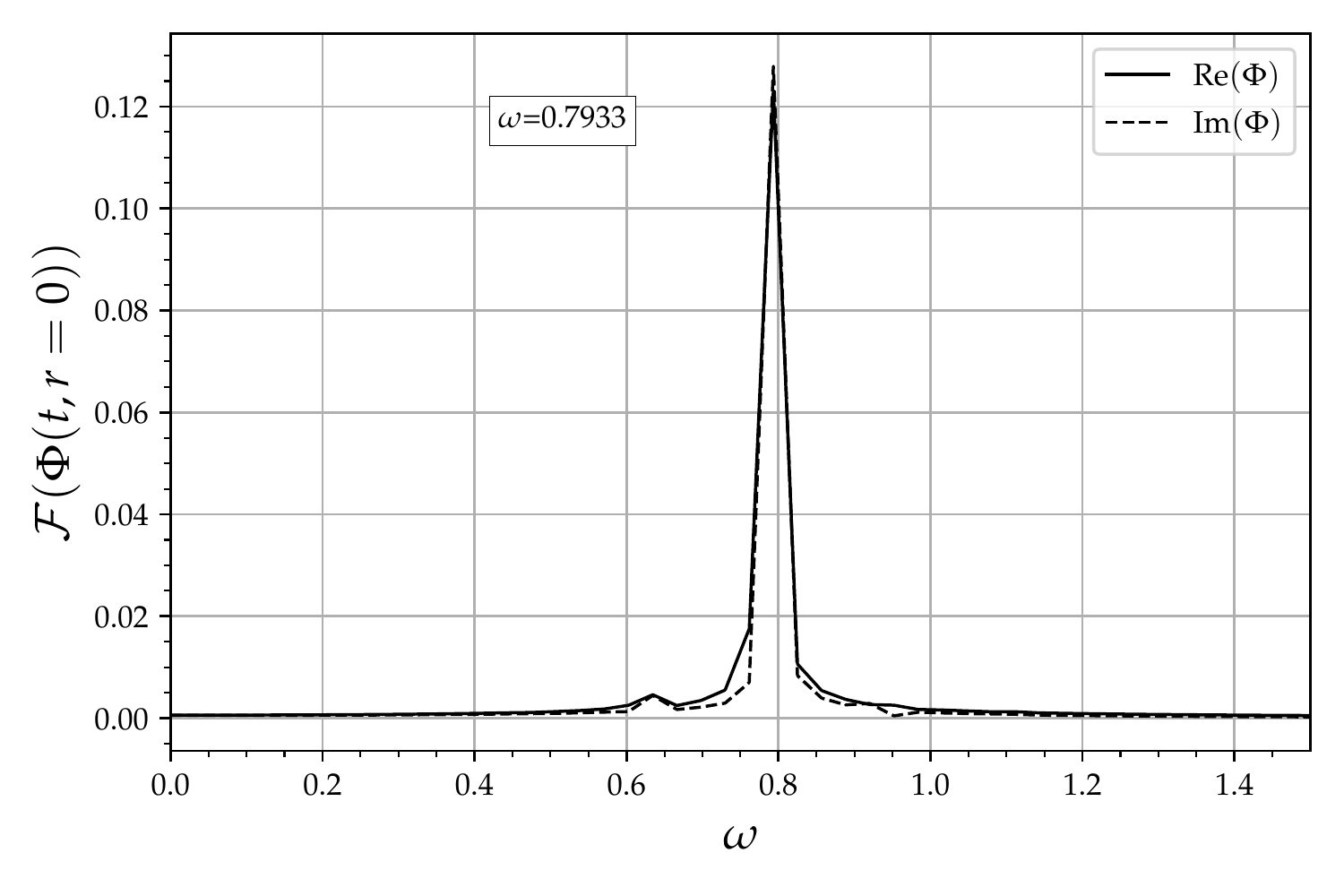}
\caption{Fourier transform of the central value of the scalar field for $\sigma=2.5$. Both the real and imaginary parts have a very narrow peak centered at a frequency $\omega=0.7933$.}
\label{fig:freq_sigma_2p5}
\end{figure}

In the time interval $25 < t < 225$, we obtain the mean value of the norm of the complex field at the origin, and then look for the $R_{99}$ of the corresponding boson star. Figure~\ref{fig:KodamaMass} shows the Kodama mass for the same subcritical evolution measured at that radius. Again, we observe an oscillation around a mean value, and the dispersion of the object after some time. We can also obtain the oscillation frequency of the scalar field by applying a Fast Fourier Transform (FFT) to the central value of its real and imaginary parts. Figure~\ref{fig:freq_sigma_2p5} plots the frequency obtained after applying the FFT. We can clearly see a very narrow peak centered at $\omega=0.7933$.

We summarize the results for all our simulations in Table~\ref{tab:all_data}, where we report the mean value of the norm of the scalar field at the origin, its oscillation frequency, and the Kodama mass of the critical solution. The uncertainty in the norm of the complex field and its mass are calculated from the standard deviation, and the uncertainty in the frequency is reported as half of the peak width in the FFT. The uncertainty in the critical exponent is obtained from the method of least squares applied to equation~\eqref{eq:tau_scaling}.

\begin{table}
\begin{tabular}{|c|c|c|c|c|}
   		\hline
   		$\sigma$ & $|\bar{\Phi}(r=0)|$        & $\bar{M}$               & $\omega$          & $\gamma$           \\ \hline
   		2.5      & 0.127$\pm$0.008 & 0.590 $\pm$ 0.003 & 0.793 $\pm$ 0.008 & 5.075 $\pm$ 0.024  \\
   		2.75     & 0.115$\pm$0.007 & 0.606 $\pm$ 0.002 & 0.804 $\pm$ 0.006 & 6.884 $\pm$ 0.022  \\
   		3        & 0.106$\pm$0.006 & 0.615 $\pm$ 0.003 & 0.813 $\pm$ 0.004 & 9.168 $\pm$ 0.023  \\
   		3.5      & 0.092$\pm$0.004 & 0.629 $\pm$ 0.002 & 0.833 $\pm$ 0.008 & 15.592 $\pm$ 0.042 \\
   		4        & 0.086$\pm$0.004 & 0.631 $\pm$ 0.002 & 0.842 $\pm$ 0.006 & 20.887 $\pm$ 0.061 \\
   		5        & 0.092$\pm$0.003 & 0.628 $\pm$ 0.002 & 0.832 $\pm$ 0.008 & 15.494 $\pm$ 0.023 \\
   		6        & 0.098$\pm$0.004 & 0.623 $\pm$ 0.003 & 0.826 $\pm$ 0.009 & 11.567 $\pm$ 0.035 \\
   		7        & 0.104$\pm$0.004 & 0.621 $\pm$ 0.003 & 0.820 $\pm$ 0.011 & 9.572 $\pm$ 0.022  \\
   		8        & 0.108$\pm$0.005 & 0.613 $\pm$ 0.004 & 0.813 $\pm$ 0.012 & 8.423 $\pm$ 0.023  \\
   		9        & 0.112$\pm$0.006 & 0.609 $\pm$ 0.004 & 0.810 $\pm$ 0.014 & 7.608 $\pm$ 0.016  \\
   		10       & 0.115$\pm$0.007 & 0.605 $\pm$ 0.005 & 0.806 $\pm$ 0.006 & 7.051 $\pm$ 0.024  \\ 
   		\hline
\end{tabular}
\caption{Summary of our numerical results for the critical solutions. Since $\Phi(r=0)$ and $M$ have an oscillatory behavior, we report the mean value with an uncertainty given by the standard deviation. The frequency is obtained using a FFT applied to the real and imaginary parts of the field at the origin. The critical exponent $\gamma$ is calculated using a least squares fit to eq.~\eqref{eq:tau_scaling}. We notice that as $\sigma$ increases, the mass of the critical solution first approaches the highest possible value for the mass for a boson star $M \sim 0.633$, reaching the maximum value for $\sigma=4.0$, while for higher values of $\sigma$ the mass decreases again.}
\label{tab:all_data}
\end{table}

\vspace{5mm}

\begin{figure}
\includegraphics[width=0.75\textwidth]{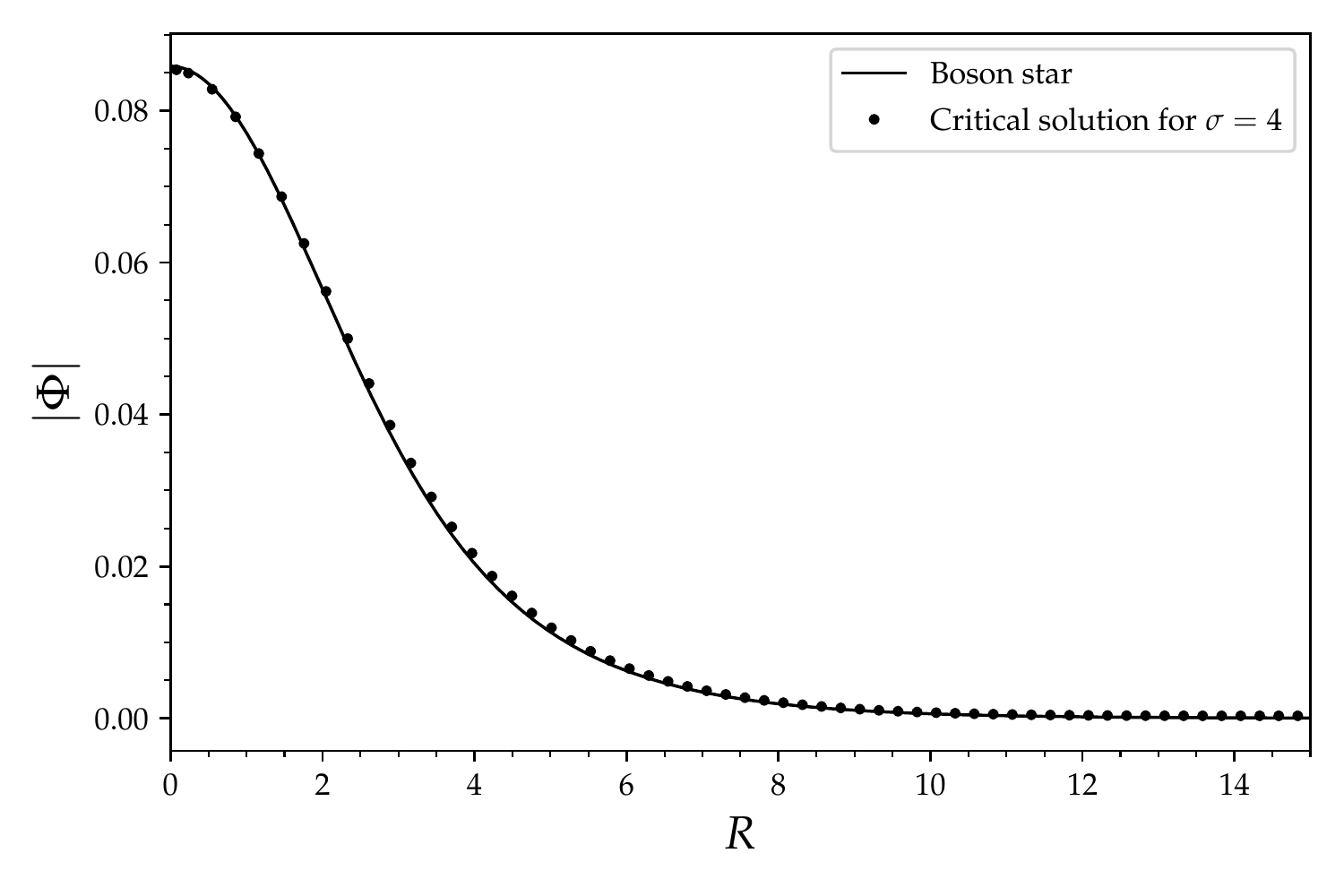}
\includegraphics[width=0.75\textwidth]{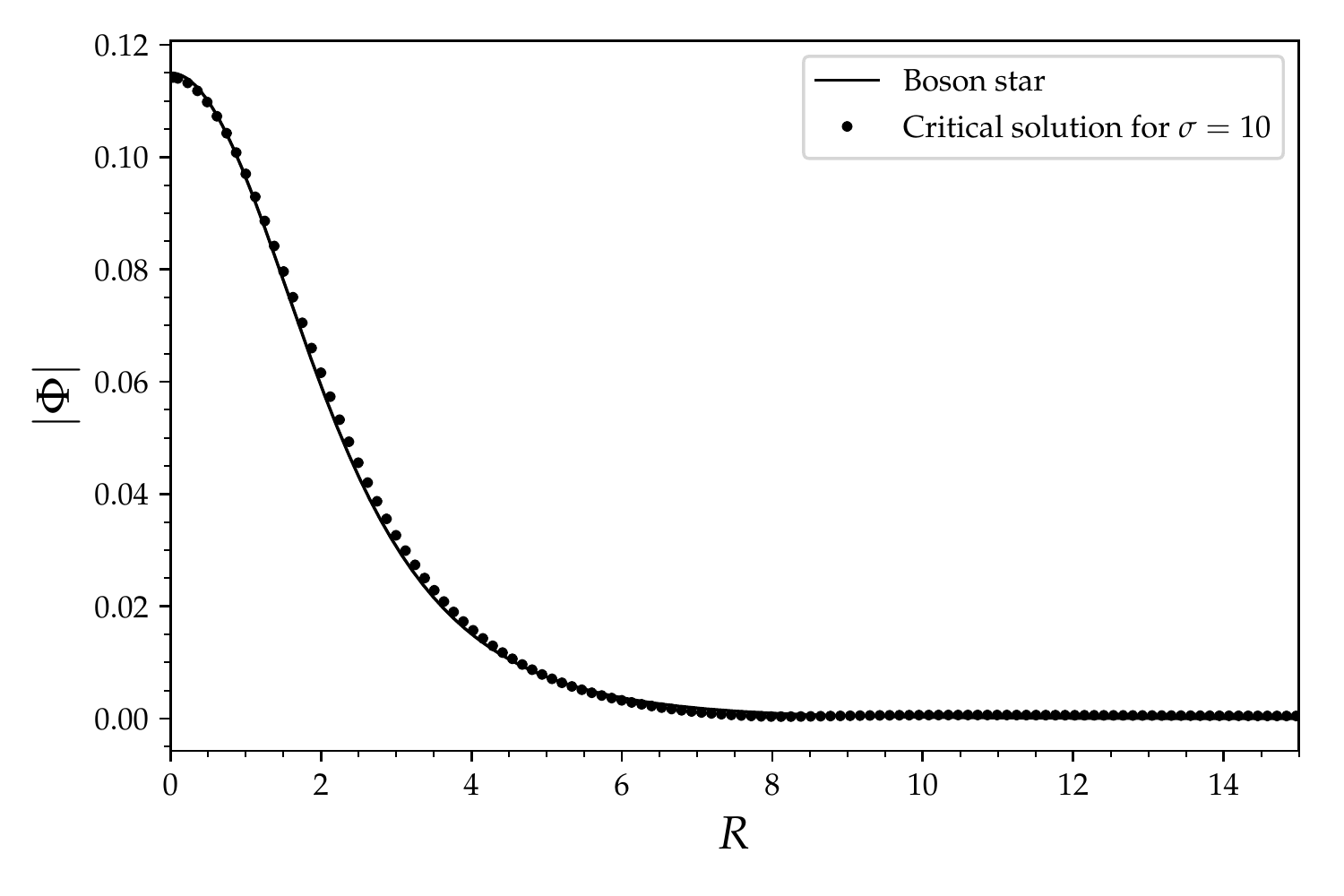}
\caption{Comparison of the norm of the complex scalar field of our critical solutions with that of unstable boson stars with the same amplitude, for the cases $\sigma=4$ (top pannel) and $\sigma=10$ (bottom panel). The dots represents the critical solutions, and the solid lines the corresponding boson stars.}
\label{fig:BS-s4-10}
\end{figure}

We now turn to the question of whether our critical solutions do indeed correspond to unstable boson stars as was found by Hawley and Choptuik in~\cite{PhysRevD.62.104024}, even if our initial data is very different. As a first comparison, Figure \ref{fig:BS-s4-10} shows the norm of the complex scalar field for our critical solution as a function of areal radius for the cases with $\sigma=4$ (top pannel) and $\sigma=10$ (bottom panel), superimposed with the norm of the complex scalar field for an unstable boson star with the same amplitude. We notice that the critical solutions obtained have no nodes in the field, so the corresponding boson star is in its ground state. We can clearly see that the profiles of our critical solutions follow very closely the expected profile for the boson stars.

\begin{figure}
\includegraphics[width=0.75\textwidth]{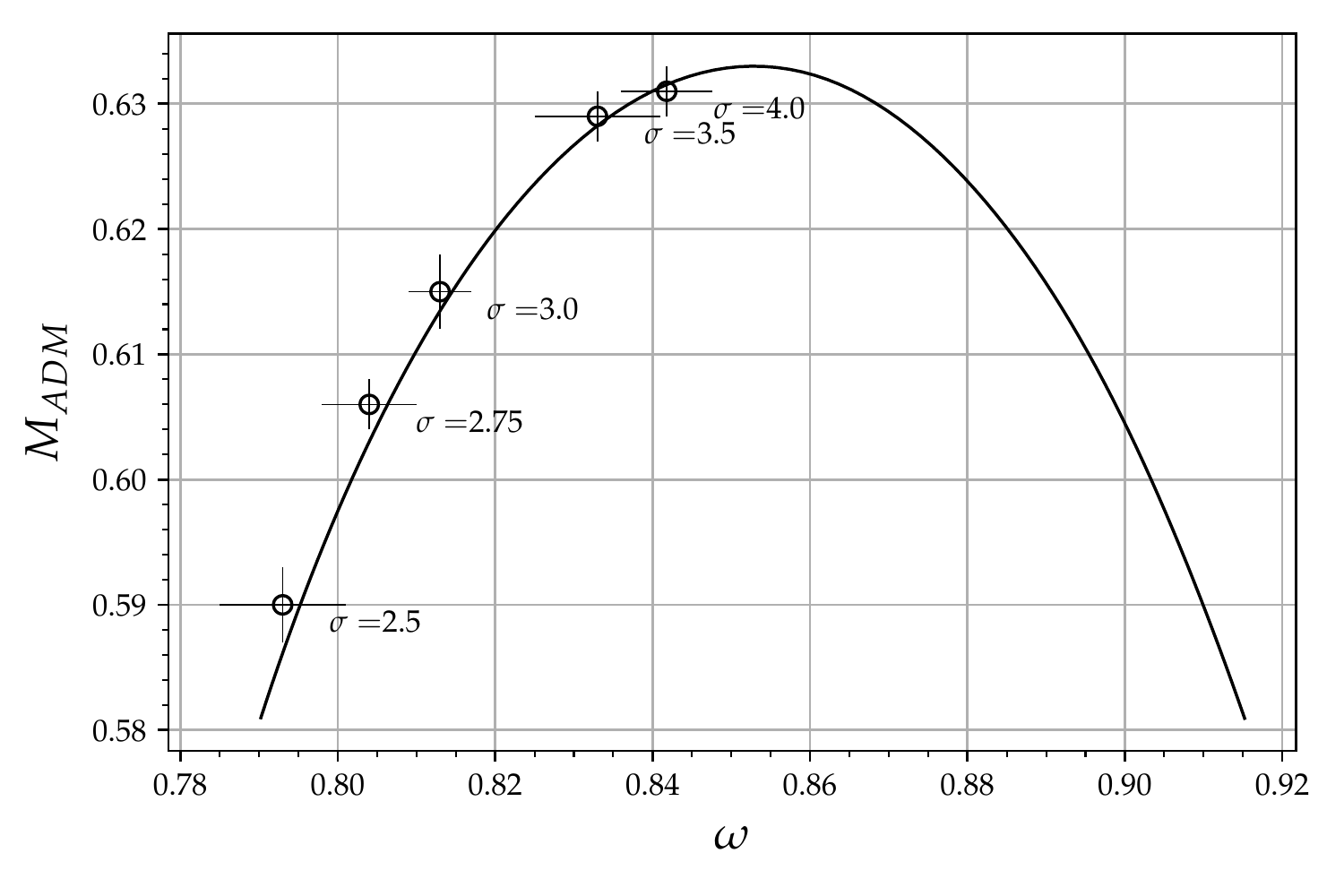}
\includegraphics[width=0.75\textwidth]{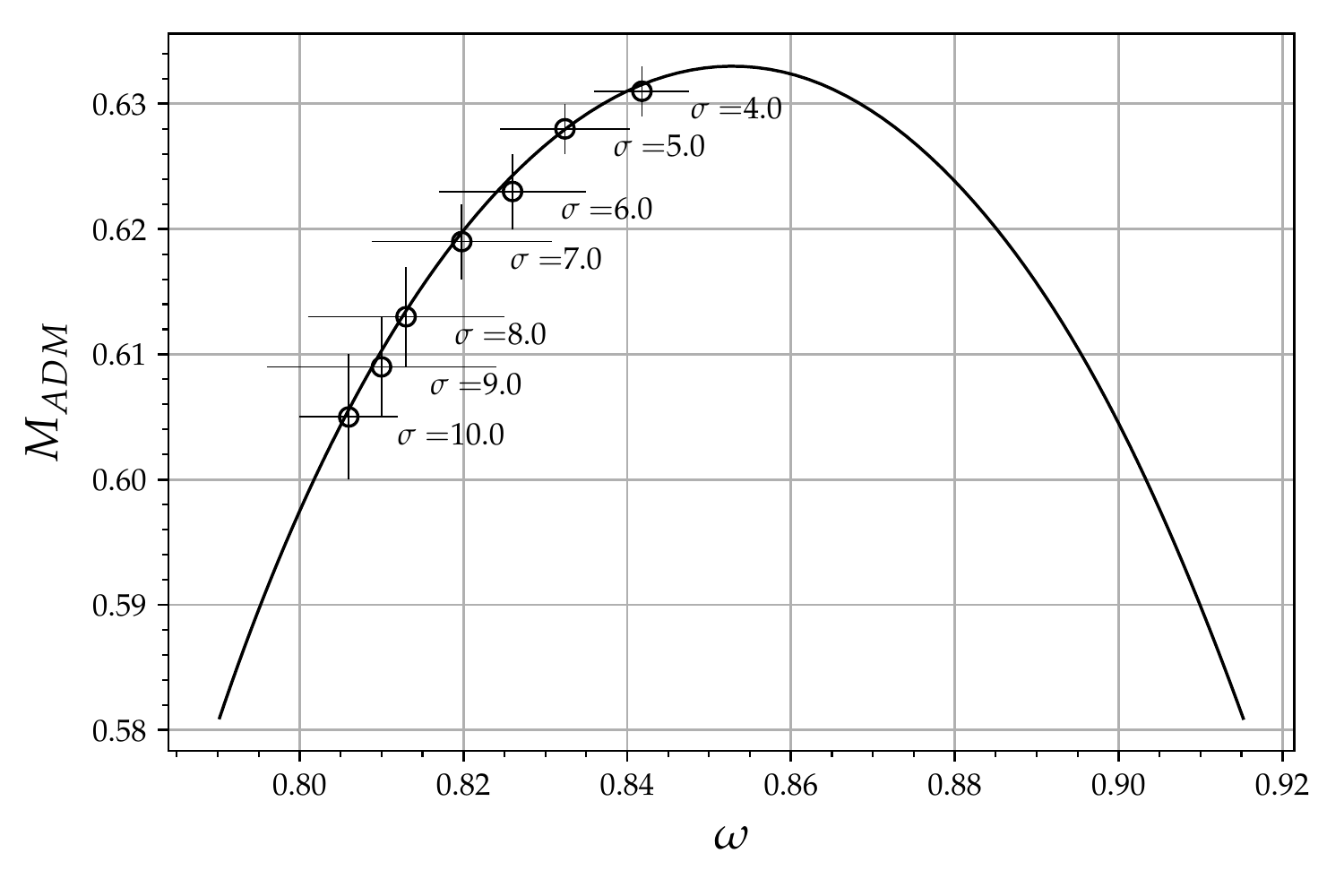}
\caption{Mass and oscillation frequency for our critical solutions compared with the curve for stationary boson star solutions. Circles corresponds to the specific value of $\sigma$ and the solid line to the known values for stationary boson stars. Top panel: Critical solutions for $\sigma \leq 4$. Bottom panel: Critical solutions for $\sigma \geq 4$. We observe that the maximum value for the mass is reached for $\sigma$ such that $3.5 < \sigma < 5$.}
\label{fig:mass_freq}
\end{figure}

\begin{figure}
\includegraphics[width=0.75\textwidth]{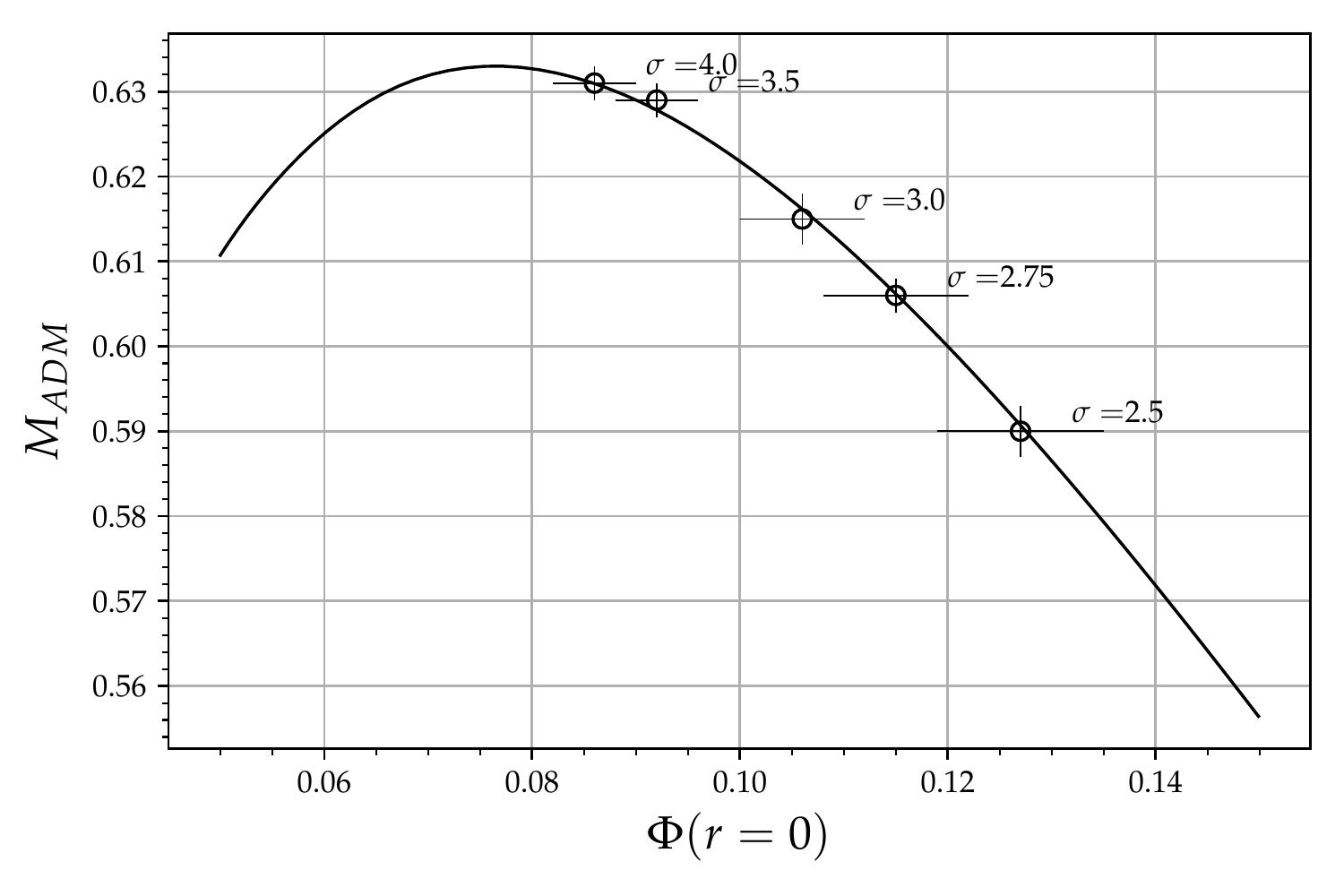}
\includegraphics[width=0.75\textwidth]{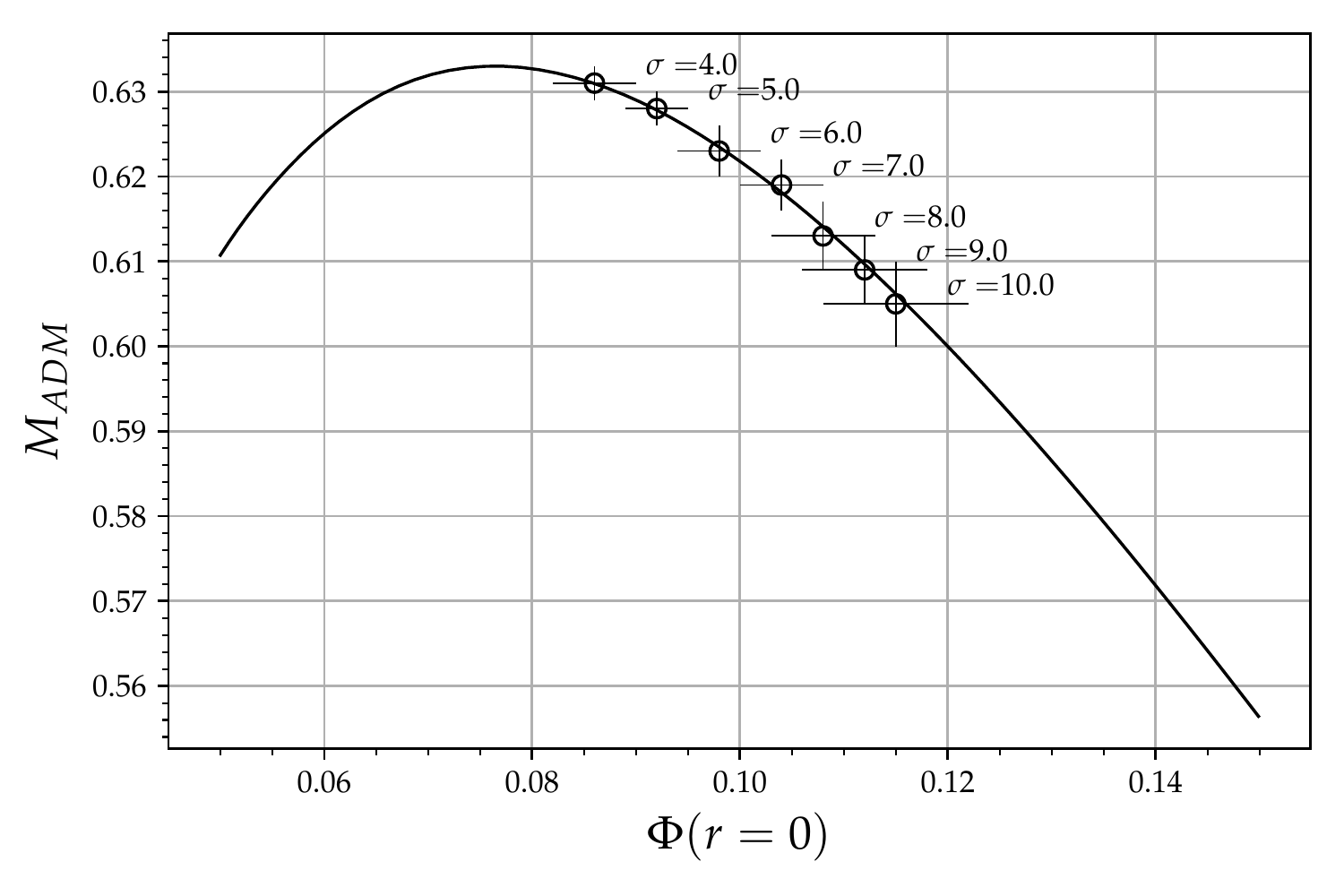}
\caption{Mass and scalar field norm at the origin for our critical solutions compared with the curve for stationary boson star solutions. Circles corresponds to the specific value of $\sigma$ and the solid line to the known values for stationary boson stars. Top panel: Critical solutions for $\sigma \leq 4$. Bottom panel: Critical solutions for $\sigma \geq 4$.}
\label{fig:mass_phi}
\end{figure}

\begin{figure}
\includegraphics[width=0.75\textwidth]{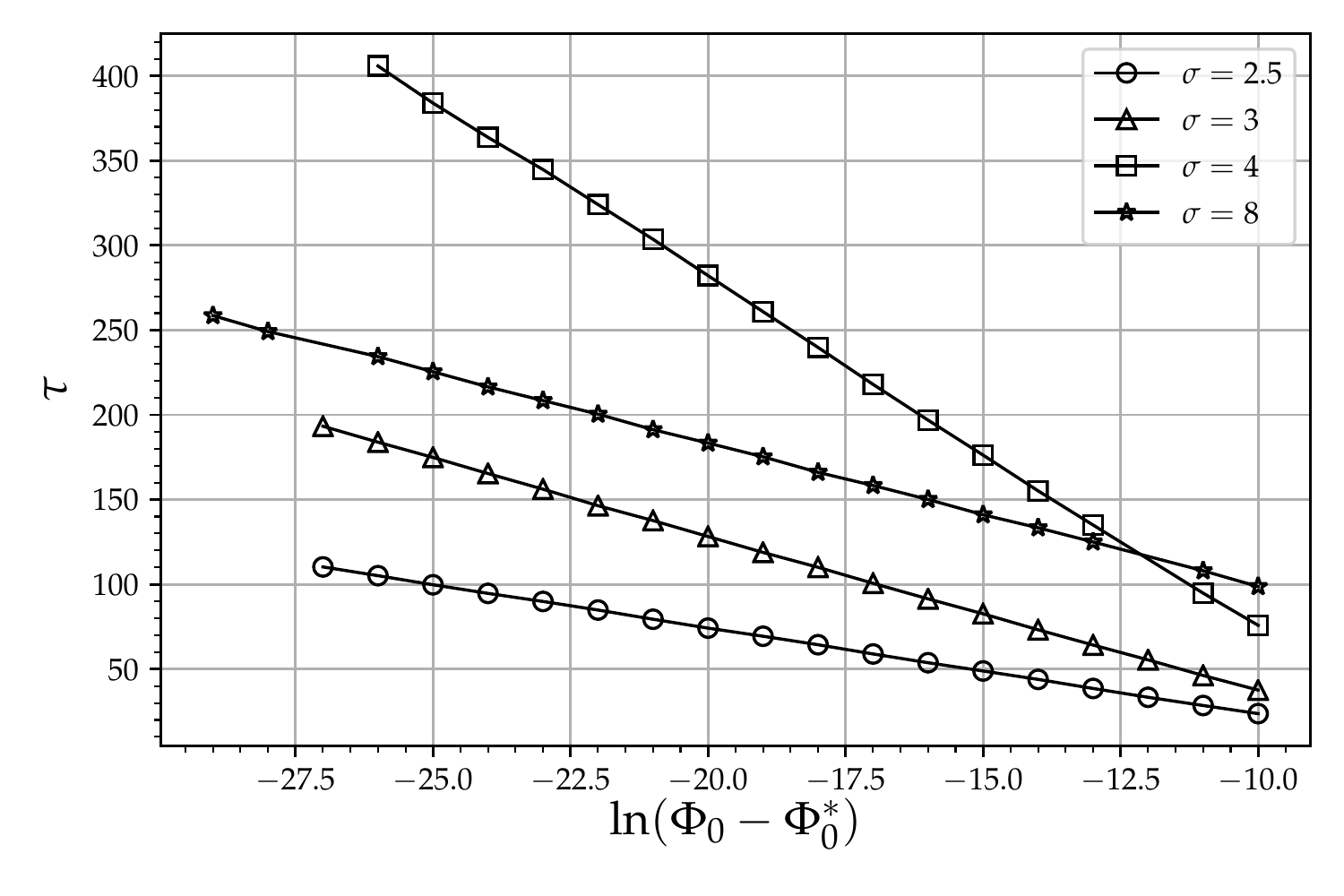}
\caption{Scaling of the lifetime of near critical solutions for different values of the gaussian width $\sigma$. With the values of $\sigma$ tested, the maximum value of the critical exponent $\gamma$ is reached at $\sigma=4$. }
\label{fig:scaling_tau}
\end{figure}

Next, in Figure~\ref{fig:mass_freq} we show a plot of the mass vs. the frequency of oscillation for our critical solutions, corresponding to the data in table~\ref{tab:all_data}, compared to the same plot for boson star solutions (solid lines).  We separate the data into two plots to make more evident the fact that the mass of the critical solution first increases with $\sigma$ up to $\sigma=4$, and then decreases again with higher values of $\sigma$. Figure~\ref{fig:mass_phi} shows a similar plot but now of the mass vs. the central value of the norm of the scalar field. As can be seen in the plots, our critical solutions fall directly in the line for stationary boson stars.  Moreover, they are all to the left of the maximum mass in Figure~\ref{fig:mass_freq}, and to the right of the maximum in Figure~\ref{fig:mass_phi}, which correspond to the unstable branch for boson stars.  The critical solution for $\sigma=4$ is almost at the maximum mass. Figure~\ref{fig:scaling_tau} shows the scaling of $\tau$, the lifetime of the near critical solutions for the different values of $\sigma$. We can see that the scaling shows good agreement with Eq.~\eqref{eq:tau_scaling}, but with different critical exponents for the different values of $\sigma$.

Since the ADM mass of the initial pulse increases monotonically relative to $\sigma$, we could have expected that the mass of the critical solution approaches asymptotically the maximum value for boson stars $M_{ADM} \sim 0.633$, which separates the unstable from the stable regions, as $\sigma$ is increased. But as can be seen from the plots and the data of Table~\ref{tab:all_data}, instead we find that for $\sigma \lesssim 4$ the mass of the critical solution increases, while for higher values of $\sigma$ the mass decreases and moves away from the maximum mass value.  Our data indicates that the maximum possible mass for a boson star will probably be attained for $\sigma $ between $3.5< \sigma <5$. This behavior is also reflected in the values of critical exponent $\gamma$, which also reaches its maximum value between $3.5 < \sigma < 5$.

\begin{figure}
\includegraphics[width=0.75\textwidth]{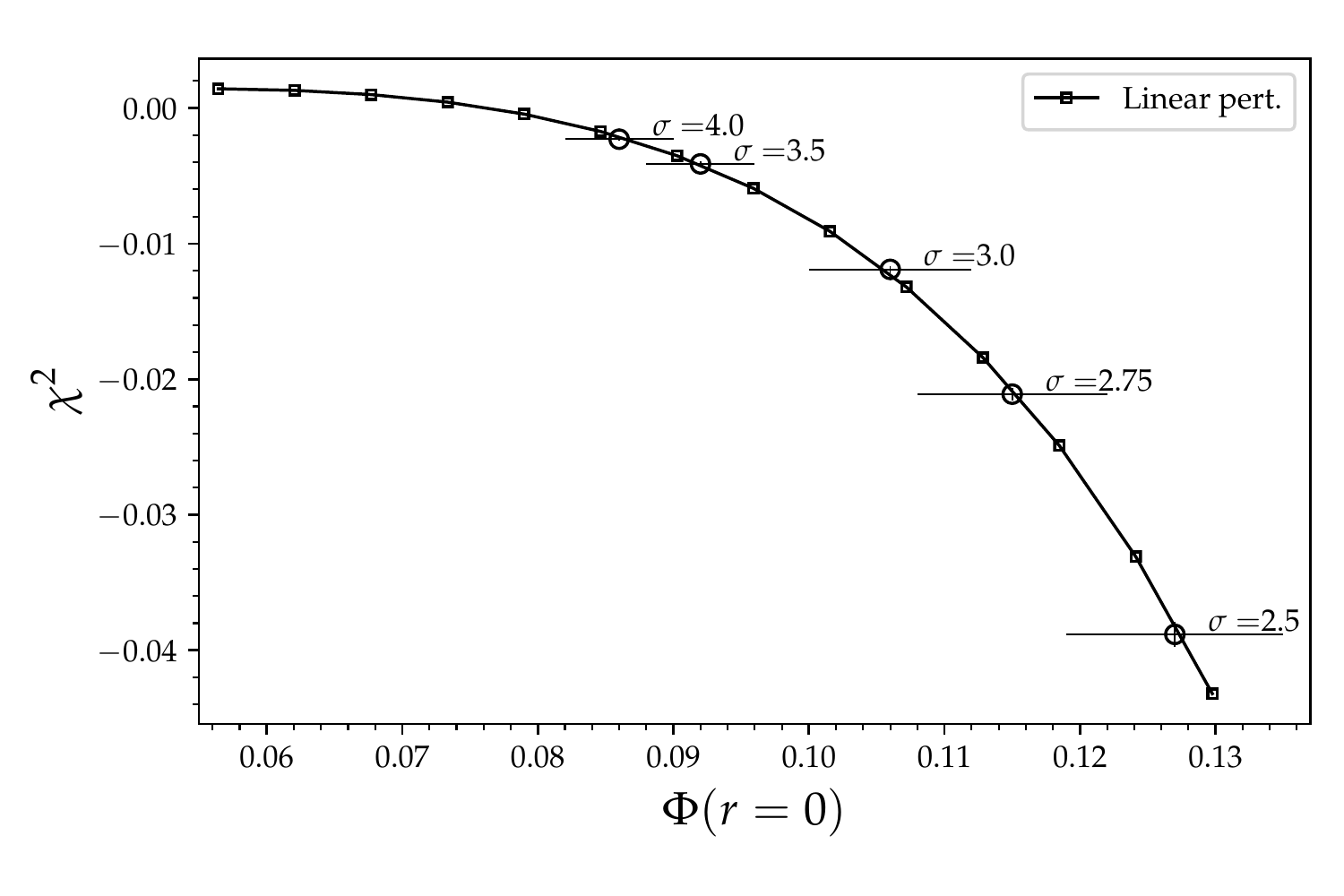}
\includegraphics[width=0.75\textwidth]{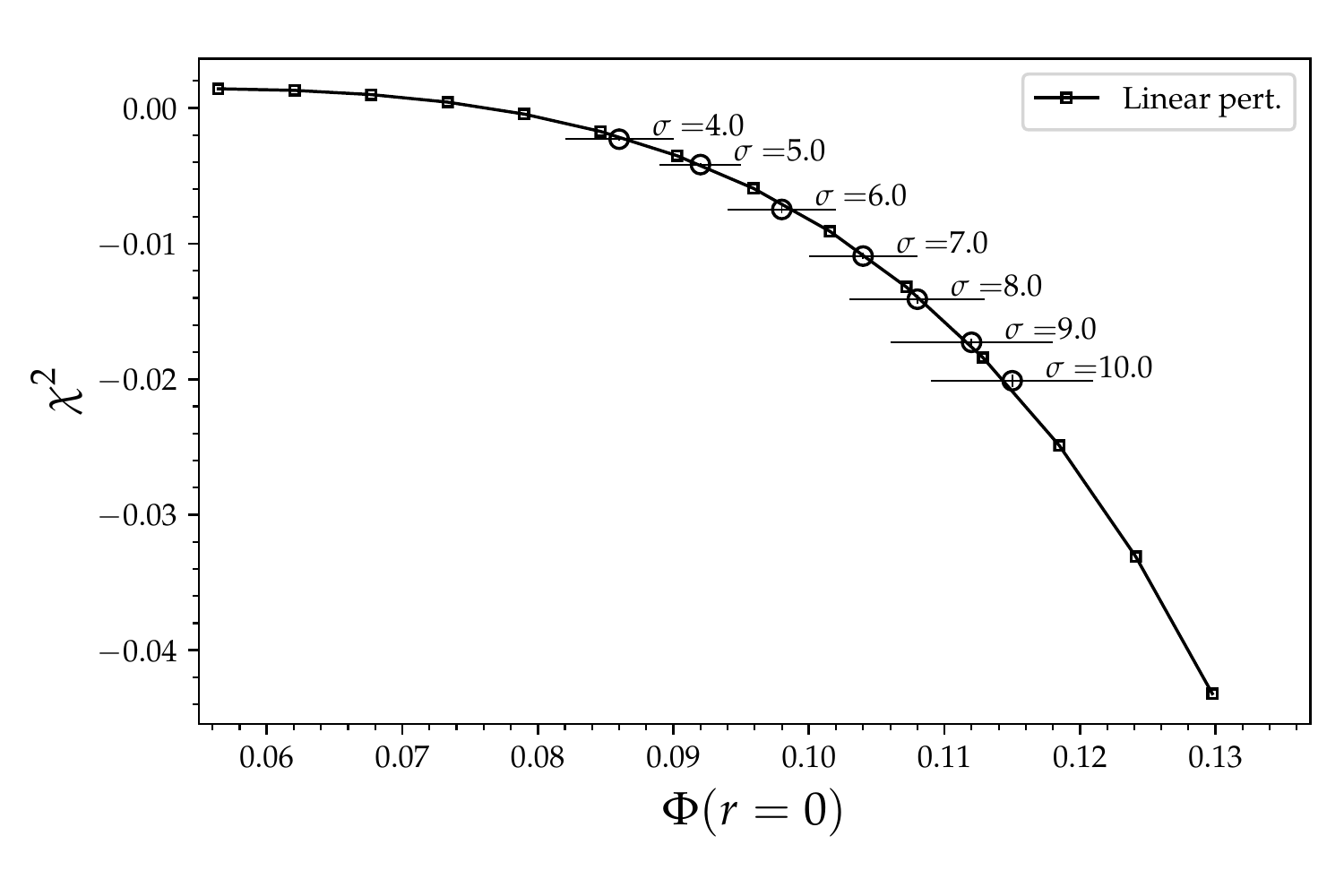}
\caption {Comparison of the square of Lyapunov exponent for unstable modes of boson stars. Circles correspond to the $1/\gamma^2$ for the specific values of $\sigma$, while the solid lines correspond to the Lyapunov exponents obtaned from a linear perturbation analysis of boson star solutions. Top panel: for $\sigma \leq 4$. Bottom panel: $\sigma \geq 4$.}
\label{fig:lyapunov}
\end{figure}

Finally, for boson stars the imaginary part of Lyapunov exponent $\lambda$ can be related to the critical exponent $\gamma$, by \mbox{$\text{Im}(\chi) = 1/\gamma$} (details about the procedure to obtain the Lyapunov exponents can be consulted in~\cite{PhysRevD.62.104024}). Figure~\ref{fig:lyapunov} compares the square of the Lyapunov exponent for boson stars obtained through a linear perturbation analysis (data provided by A.~Bernal~\cite{BernalA_2021}), with the critical exponents $1/\gamma^2$ measured in our simulations. We can see an excellent agreement between both data sets.

\section{Conclusions}

We performed numerical simulations of a massive complex scalar field using a numerical code adapted to spherical symmetry in order to study critical gravitational collapse. Our initial conditions for the complex field are somewhat similar to the harmonic boson star ansatz, but crucially they do not correspond to a stationary boson star solution.

We find that, depending on the width of initial data, the critical collapse behaves in two very different ways. For $\sigma \leq 0.5$ we can measure the 4D Ricci scaling, which is indicative of type II critical collapse. We obtain values for the critical exponent $\gamma = 0.38 \pm 0.01$ and echoing period $\Delta = 3.4 \pm 0.1$, which are very similar to those found in the literaure for the case of a real massless scalar field. On the other hand, for $\sigma \geq 2.5$ we obtain the scaling of the lifetime of near critical solutions, which is characteristic of type I critical collapse. For type I collapse we observe that the critical exponent depends on the initial gaussian width $\sigma$: as this width increases the critical exponent reaches its highest value for $\sigma \approx 4$, while for higher values of $\sigma$ the value of the critical exponent decreases again.

In a similar way to Hawley and Choptuik~\cite{PhysRevD.62.104024}, we also find that the critical solutions obtained correspond to boson stars in the ground state in the unstable branch. We validate our results by contrasting the curves of $|\Phi(0)|$ vs. $M_{ADM}$, and $\omega$ vs $M_{ADM}$. Up to our uncertainties we find that our critical solutions do fall on the curves for unstable stationary boson stars. Also, for our simulations the maximum mass of the critical solution is obtained for $\sigma \simeq 4$, which leads us to conjecture that the maximum mass of a boson star at the boundary between the stable and unstable branches, $M \sim 0.633$,  will be attained for $\sigma$ somewhere in the range $3.5<\sigma<5.0$.  Furthermore, we also confirm that the inverse of the critical exponent $\gamma$ for our critical solutions does indeed correspond to the imaginary part of Lyapunov exponents for unstable boson stars obtained through a linear perturbation analysis.

One final comment about the transition from type I critical collapse to type II. Since we have obtained the two different behaviors by varying the value of $\sigma$, we can in principle study the transition between both types of collapse by concentrating in the region $0.5 < \sigma < 2.5$. On the other hand, as can be seen from our plots, for type I collapse as $\sigma$ decreases the critical solution moves further into the unstable branch for boson stars. This raises the question as to how far down this branch we can go before transitioning to type II critical collapse. We will leave a more detailed study of this question for a future study.


\acknowledgments

This work was partially supported by CONACyT Network Projects No. 376127 and No. 304001. EJ was also supported by a CONACyT National Graduate Grant.


\bibliography{references}

\begin{thebibliography}{36}
\expandafter\ifx\csname natexlab\endcsname\relax\def\natexlab#1{#1}\fi
\expandafter\ifx\csname bibnamefont\endcsname\relax
  \def\bibnamefont#1{#1}\fi
\expandafter\ifx\csname bibfnamefont\endcsname\relax
  \def\bibfnamefont#1{#1}\fi
\expandafter\ifx\csname citenamefont\endcsname\relax
  \def\citenamefont#1{#1}\fi
\expandafter\ifx\csname url\endcsname\relax
  \def\url#1{\texttt{#1}}\fi
\expandafter\ifx\csname urlprefix\endcsname\relax\def\urlprefix{URL }\fi
\providecommand{\bibinfo}[2]{#2}
\providecommand{\eprint}[2][]{\url{#2}}

\bibitem[{\citenamefont{Choptuik}(1993)}]{PhysRevLett.70.9}
\bibinfo{author}{\bibfnamefont{M.~W.} \bibnamefont{Choptuik}},
  \bibinfo{journal}{Phys. Rev. Lett.} \textbf{\bibinfo{volume}{70}},
  \bibinfo{pages}{9} (\bibinfo{year}{1993}),
  \urlprefix\url{https://link.aps.org/doi/10.1103/PhysRevLett.70.9}.

\bibitem[{\citenamefont{Rinne}(2020)}]{Rinne:2020asi}
\bibinfo{author}{\bibfnamefont{O.}~\bibnamefont{Rinne}}, \bibinfo{journal}{Gen.
  Rel. Grav.} \textbf{\bibinfo{volume}{52}}, \bibinfo{pages}{117}
  (\bibinfo{year}{2020}), \eprint{2008.12726}.

\bibitem[{\citenamefont{Gundlach et~al.}(1994)\citenamefont{Gundlach, Price,
  and Pullin}}]{PhysRevD.49.890}
\bibinfo{author}{\bibfnamefont{C.}~\bibnamefont{Gundlach}},
  \bibinfo{author}{\bibfnamefont{R.~H.} \bibnamefont{Price}}, \bibnamefont{and}
  \bibinfo{author}{\bibfnamefont{J.}~\bibnamefont{Pullin}},
  \bibinfo{journal}{Phys. Rev. D} \textbf{\bibinfo{volume}{49}},
  \bibinfo{pages}{890} (\bibinfo{year}{1994}),
  \urlprefix\url{https://link.aps.org/doi/10.1103/PhysRevD.49.890}.

\bibitem[{\citenamefont{Garfinkle}(1995)}]{PhysRevD.51.5558}
\bibinfo{author}{\bibfnamefont{D.}~\bibnamefont{Garfinkle}},
  \bibinfo{journal}{Phys. Rev. D} \textbf{\bibinfo{volume}{51}},
  \bibinfo{pages}{5558} (\bibinfo{year}{1995}),
  \urlprefix\url{https://link.aps.org/doi/10.1103/PhysRevD.51.5558}.

\bibitem[{\citenamefont{Hamad{\'{e}} and Stewart}(1996)}]{Hamad__1996}
\bibinfo{author}{\bibfnamefont{R.~S.} \bibnamefont{Hamad{\'{e}}}}
  \bibnamefont{and} \bibinfo{author}{\bibfnamefont{J.~M.}
  \bibnamefont{Stewart}}, \bibinfo{journal}{Classical and Quantum Gravity}
  \textbf{\bibinfo{volume}{13}}, \bibinfo{pages}{497} (\bibinfo{year}{1996}),
  \urlprefix\url{https://doi.org/10.1088/0264-9381/13/3/014}.

\bibitem[{\citenamefont{Akbarian and Choptuik}(2015)}]{PhysRevD.92.084037}
\bibinfo{author}{\bibfnamefont{A.}~\bibnamefont{Akbarian}} \bibnamefont{and}
  \bibinfo{author}{\bibfnamefont{M.~W.} \bibnamefont{Choptuik}},
  \bibinfo{journal}{Phys. Rev. D} \textbf{\bibinfo{volume}{92}},
  \bibinfo{pages}{084037} (\bibinfo{year}{2015}),
  \urlprefix\url{https://link.aps.org/doi/10.1103/PhysRevD.92.084037}.

\bibitem[{\citenamefont{Choptuik et~al.}(1996)\citenamefont{Choptuik, Chmaj,
  and Bizo\ifmmode~\acute{n}\else \'{n}\fi{}}}]{PhysRevLett.77.424}
\bibinfo{author}{\bibfnamefont{M.~W.} \bibnamefont{Choptuik}},
  \bibinfo{author}{\bibfnamefont{T.}~\bibnamefont{Chmaj}}, \bibnamefont{and}
  \bibinfo{author}{\bibfnamefont{P.}~\bibnamefont{Bizo\ifmmode~\acute{n}\else
  \'{n}\fi{}}}, \bibinfo{journal}{Phys. Rev. Lett.}
  \textbf{\bibinfo{volume}{77}}, \bibinfo{pages}{424} (\bibinfo{year}{1996}),
  \urlprefix\url{https://link.aps.org/doi/10.1103/PhysRevLett.77.424}.

\bibitem[{\citenamefont{Brady et~al.}(1997)\citenamefont{Brady, Chambers, and
  Gon\ifmmode~\mbox{\c{c}}\else \c{c}\fi{}alves}}]{PhysRevD.56.R6057}
\bibinfo{author}{\bibfnamefont{P.~R.} \bibnamefont{Brady}},
  \bibinfo{author}{\bibfnamefont{C.~M.} \bibnamefont{Chambers}},
  \bibnamefont{and} \bibinfo{author}{\bibfnamefont{S.~M. C.~V.}
  \bibnamefont{Gon\ifmmode~\mbox{\c{c}}\else \c{c}\fi{}alves}},
  \bibinfo{journal}{Phys. Rev. D} \textbf{\bibinfo{volume}{56}},
  \bibinfo{pages}{R6057} (\bibinfo{year}{1997}),
  \urlprefix\url{https://link.aps.org/doi/10.1103/PhysRevD.56.R6057}.

\bibitem[{\citenamefont{Gundlach and Martin-Garcia}(2007)}]{Gundlach:2007gc}
\bibinfo{author}{\bibfnamefont{C.}~\bibnamefont{Gundlach}} \bibnamefont{and}
  \bibinfo{author}{\bibfnamefont{J.~M.} \bibnamefont{Martin-Garcia}},
  \bibinfo{journal}{Living Rev. Rel.} \textbf{\bibinfo{volume}{10}},
  \bibinfo{pages}{5} (\bibinfo{year}{2007}), \eprint{0711.4620}.

\bibitem[{\citenamefont{Hawley and Choptuik}(2000)}]{PhysRevD.62.104024}
\bibinfo{author}{\bibfnamefont{S.~H.} \bibnamefont{Hawley}} \bibnamefont{and}
  \bibinfo{author}{\bibfnamefont{M.~W.} \bibnamefont{Choptuik}},
  \bibinfo{journal}{Phys. Rev. D} \textbf{\bibinfo{volume}{62}},
  \bibinfo{pages}{104024} (\bibinfo{year}{2000}),
  \urlprefix\url{https://link.aps.org/doi/10.1103/PhysRevD.62.104024}.

\bibitem[{\citenamefont{Kaup}(1968)}]{PhysRev.172.1331}
\bibinfo{author}{\bibfnamefont{D.~J.} \bibnamefont{Kaup}},
  \bibinfo{journal}{Phys. Rev.} \textbf{\bibinfo{volume}{172}},
  \bibinfo{pages}{1331} (\bibinfo{year}{1968}),
  \urlprefix\url{https://link.aps.org/doi/10.1103/PhysRev.172.1331}.

\bibitem[{\citenamefont{Ruffini and Bonazzola}(1969)}]{PhysRev.187.1767}
\bibinfo{author}{\bibfnamefont{R.}~\bibnamefont{Ruffini}} \bibnamefont{and}
  \bibinfo{author}{\bibfnamefont{S.}~\bibnamefont{Bonazzola}},
  \bibinfo{journal}{Phys. Rev.} \textbf{\bibinfo{volume}{187}},
  \bibinfo{pages}{1767} (\bibinfo{year}{1969}),
  \urlprefix\url{https://link.aps.org/doi/10.1103/PhysRev.187.1767}.

\bibitem[{\citenamefont{Visinelli}(2021)}]{Visinelli:2021uve}
\bibinfo{author}{\bibfnamefont{L.}~\bibnamefont{Visinelli}},
  \bibinfo{journal}{Int. J. Mod. Phys. D} \textbf{\bibinfo{volume}{30}},
  \bibinfo{pages}{2130006} (\bibinfo{year}{2021}), \eprint{2109.05481}.

\bibitem[{\citenamefont{Liebling and Palenzuela}(2012)}]{Liebling:2012fv}
\bibinfo{author}{\bibfnamefont{S.~L.} \bibnamefont{Liebling}} \bibnamefont{and}
  \bibinfo{author}{\bibfnamefont{C.}~\bibnamefont{Palenzuela}},
  \bibinfo{journal}{Living Rev. Rel.} \textbf{\bibinfo{volume}{15}},
  \bibinfo{pages}{6} (\bibinfo{year}{2012}), \eprint{1202.5809}.

\bibitem[{\citenamefont{Seidel and Suen}(1990)}]{PhysRevD.42.384}
\bibinfo{author}{\bibfnamefont{E.}~\bibnamefont{Seidel}} \bibnamefont{and}
  \bibinfo{author}{\bibfnamefont{W.-M.} \bibnamefont{Suen}},
  \bibinfo{journal}{Phys. Rev. D} \textbf{\bibinfo{volume}{42}},
  \bibinfo{pages}{384} (\bibinfo{year}{1990}),
  \urlprefix\url{https://link.aps.org/doi/10.1103/PhysRevD.42.384}.

\bibitem[{\citenamefont{Gleiser and Watkins}(1989)}]{GLEISER1989733}
\bibinfo{author}{\bibfnamefont{M.}~\bibnamefont{Gleiser}} \bibnamefont{and}
  \bibinfo{author}{\bibfnamefont{R.}~\bibnamefont{Watkins}},
  \bibinfo{journal}{Nuclear Physics B} \textbf{\bibinfo{volume}{319}},
  \bibinfo{pages}{733} (\bibinfo{year}{1989}), ISSN \bibinfo{issn}{0550-3213},
  \urlprefix\url{https://www.sciencedirect.com/science/article/pii/0550321389906275}.

\bibitem[{\citenamefont{Shibata and Nakamura}(1995)}]{PhysRevD.52.5428}
\bibinfo{author}{\bibfnamefont{M.}~\bibnamefont{Shibata}} \bibnamefont{and}
  \bibinfo{author}{\bibfnamefont{T.}~\bibnamefont{Nakamura}},
  \bibinfo{journal}{Phys. Rev. D} \textbf{\bibinfo{volume}{52}},
  \bibinfo{pages}{5428} (\bibinfo{year}{1995}),
  \urlprefix\url{https://link.aps.org/doi/10.1103/PhysRevD.52.5428}.

\bibitem[{\citenamefont{Baumgarte and Shapiro}(1998)}]{PhysRevD.59.024007}
\bibinfo{author}{\bibfnamefont{T.~W.} \bibnamefont{Baumgarte}}
  \bibnamefont{and} \bibinfo{author}{\bibfnamefont{S.~L.}
  \bibnamefont{Shapiro}}, \bibinfo{journal}{Phys. Rev. D}
  \textbf{\bibinfo{volume}{59}}, \bibinfo{pages}{024007}
  (\bibinfo{year}{1998}),
  \urlprefix\url{https://link.aps.org/doi/10.1103/PhysRevD.59.024007}.

\bibitem[{\citenamefont{Sarbach et~al.}(2002)\citenamefont{Sarbach, Calabrese,
  Pullin, and Tiglio}}]{PhysRevD.66.064002}
\bibinfo{author}{\bibfnamefont{O.}~\bibnamefont{Sarbach}},
  \bibinfo{author}{\bibfnamefont{G.}~\bibnamefont{Calabrese}},
  \bibinfo{author}{\bibfnamefont{J.}~\bibnamefont{Pullin}}, \bibnamefont{and}
  \bibinfo{author}{\bibfnamefont{M.}~\bibnamefont{Tiglio}},
  \bibinfo{journal}{Phys. Rev. D} \textbf{\bibinfo{volume}{66}},
  \bibinfo{pages}{064002} (\bibinfo{year}{2002}),
  \urlprefix\url{https://link.aps.org/doi/10.1103/PhysRevD.66.064002}.

\bibitem[{\citenamefont{Brown}(2009)}]{PhysRevD.79.104029}
\bibinfo{author}{\bibfnamefont{J.~D.} \bibnamefont{Brown}},
  \bibinfo{journal}{Phys. Rev. D} \textbf{\bibinfo{volume}{79}},
  \bibinfo{pages}{104029} (\bibinfo{year}{2009}),
  \urlprefix\url{https://link.aps.org/doi/10.1103/PhysRevD.79.104029}.

\bibitem[{\citenamefont{Alcubierre and Mendez}(2011)}]{Alcubierre:2011pkc}
\bibinfo{author}{\bibfnamefont{M.}~\bibnamefont{Alcubierre}} \bibnamefont{and}
  \bibinfo{author}{\bibfnamefont{M.~D.} \bibnamefont{Mendez}},
  \bibinfo{journal}{Gen. Rel. Grav.} \textbf{\bibinfo{volume}{43}},
  \bibinfo{pages}{2769} (\bibinfo{year}{2011}), \eprint{1010.4013}.

\bibitem[{\citenamefont{Alcubierre et~al.}(2019)\citenamefont{Alcubierre,
  Barranco, Bernal, Degollado, Diez-Tejedor, Megevand, N\'u\~nez, and
  Sarbach}}]{Alcubierre:2019qnh}
\bibinfo{author}{\bibfnamefont{M.}~\bibnamefont{Alcubierre}},
  \bibinfo{author}{\bibfnamefont{J.}~\bibnamefont{Barranco}},
  \bibinfo{author}{\bibfnamefont{A.}~\bibnamefont{Bernal}},
  \bibinfo{author}{\bibfnamefont{J.~C.} \bibnamefont{Degollado}},
  \bibinfo{author}{\bibfnamefont{A.}~\bibnamefont{Diez-Tejedor}},
  \bibinfo{author}{\bibfnamefont{M.}~\bibnamefont{Megevand}},
  \bibinfo{author}{\bibfnamefont{D.}~\bibnamefont{N\'u\~nez}},
  \bibnamefont{and} \bibinfo{author}{\bibfnamefont{O.}~\bibnamefont{Sarbach}},
  \bibinfo{journal}{Class. Quant. Grav.} \textbf{\bibinfo{volume}{36}},
  \bibinfo{pages}{215013} (\bibinfo{year}{2019}), \eprint{1906.08959}.

\bibitem[{\citenamefont{Degollado et~al.}(2020)\citenamefont{Degollado,
  Salgado, and Alcubierre}}]{Degollado:2020lsa}
\bibinfo{author}{\bibfnamefont{J.~C.} \bibnamefont{Degollado}},
  \bibinfo{author}{\bibfnamefont{M.}~\bibnamefont{Salgado}}, \bibnamefont{and}
  \bibinfo{author}{\bibfnamefont{M.}~\bibnamefont{Alcubierre}},
  \bibinfo{journal}{Phys. Lett. B} \textbf{\bibinfo{volume}{808}},
  \bibinfo{pages}{135666} (\bibinfo{year}{2020}), \eprint{2008.10683}.

\bibitem[{\citenamefont{Alcubierre}(2008)}]{Alcubierre:1138167}
\bibinfo{author}{\bibfnamefont{M.}~\bibnamefont{Alcubierre}},
  \emph{\bibinfo{title}{{Introduction to 3+1 numerical relativity}}},
  International series of monographs on physics (\bibinfo{publisher}{Oxford
  Univ. Press}, \bibinfo{address}{Oxford}, \bibinfo{year}{2008}),
  \urlprefix\url{https://cds.cern.ch/record/1138167}.

\bibitem[{\citenamefont{Kodama}(1980)}]{Kodama:1979vn}
\bibinfo{author}{\bibfnamefont{H.}~\bibnamefont{Kodama}},
  \bibinfo{journal}{Prog. Theor. Phys.} \textbf{\bibinfo{volume}{63}},
  \bibinfo{pages}{1217} (\bibinfo{year}{1980}).

\bibitem[{\citenamefont{Harada et~al.}(2015)\citenamefont{Harada, Yoo, Nakama,
  and Koga}}]{PhysRevD.91.084057}
\bibinfo{author}{\bibfnamefont{T.}~\bibnamefont{Harada}},
  \bibinfo{author}{\bibfnamefont{C.-M.} \bibnamefont{Yoo}},
  \bibinfo{author}{\bibfnamefont{T.}~\bibnamefont{Nakama}}, \bibnamefont{and}
  \bibinfo{author}{\bibfnamefont{Y.}~\bibnamefont{Koga}},
  \bibinfo{journal}{Phys. Rev. D} \textbf{\bibinfo{volume}{91}},
  \bibinfo{pages}{084057} (\bibinfo{year}{2015}),
  \urlprefix\url{https://link.aps.org/doi/10.1103/PhysRevD.91.084057}.

\bibitem[{\citenamefont{Ikeda et~al.}(2017)\citenamefont{Ikeda, Yoo, and
  Cardoso}}]{PhysRevD.96.064047}
\bibinfo{author}{\bibfnamefont{T.}~\bibnamefont{Ikeda}},
  \bibinfo{author}{\bibfnamefont{C.-M.} \bibnamefont{Yoo}}, \bibnamefont{and}
  \bibinfo{author}{\bibfnamefont{V.}~\bibnamefont{Cardoso}},
  \bibinfo{journal}{Phys. Rev. D} \textbf{\bibinfo{volume}{96}},
  \bibinfo{pages}{064047} (\bibinfo{year}{2017}),
  \urlprefix\url{https://link.aps.org/doi/10.1103/PhysRevD.96.064047}.

\bibitem[{\citenamefont{Racz}(2006)}]{Racz:2005pm}
\bibinfo{author}{\bibfnamefont{I.}~\bibnamefont{Racz}},
  \bibinfo{journal}{Class. Quant. Grav.} \textbf{\bibinfo{volume}{23}},
  \bibinfo{pages}{115} (\bibinfo{year}{2006}), \eprint{gr-qc/0511052}.

\bibitem[{\citenamefont{Alcubierre et~al.}(2018)\citenamefont{Alcubierre,
  Barranco, Bernal, Degollado, Diez-Tejedor, Megevand, Nunez, and
  Sarbach}}]{Alcubierre:2018ahf}
\bibinfo{author}{\bibfnamefont{M.}~\bibnamefont{Alcubierre}},
  \bibinfo{author}{\bibfnamefont{J.}~\bibnamefont{Barranco}},
  \bibinfo{author}{\bibfnamefont{A.}~\bibnamefont{Bernal}},
  \bibinfo{author}{\bibfnamefont{J.~C.} \bibnamefont{Degollado}},
  \bibinfo{author}{\bibfnamefont{A.}~\bibnamefont{Diez-Tejedor}},
  \bibinfo{author}{\bibfnamefont{M.}~\bibnamefont{Megevand}},
  \bibinfo{author}{\bibfnamefont{D.}~\bibnamefont{Nunez}}, \bibnamefont{and}
  \bibinfo{author}{\bibfnamefont{O.}~\bibnamefont{Sarbach}},
  \bibinfo{journal}{Class. Quant. Grav.} \textbf{\bibinfo{volume}{35}},
  \bibinfo{pages}{19LT01} (\bibinfo{year}{2018}), \eprint{1805.11488}.

\bibitem[{\citenamefont{Hod and Piran}(1997)}]{PhysRevD.55.R440}
\bibinfo{author}{\bibfnamefont{S.}~\bibnamefont{Hod}} \bibnamefont{and}
  \bibinfo{author}{\bibfnamefont{T.}~\bibnamefont{Piran}},
  \bibinfo{journal}{Phys. Rev. D} \textbf{\bibinfo{volume}{55}},
  \bibinfo{pages}{R440} (\bibinfo{year}{1997}),
  \urlprefix\url{https://link.aps.org/doi/10.1103/PhysRevD.55.R440}.

\bibitem[{\citenamefont{Baumgarte}(2018)}]{PhysRevD.98.084012}
\bibinfo{author}{\bibfnamefont{T.~W.} \bibnamefont{Baumgarte}},
  \bibinfo{journal}{Phys. Rev. D} \textbf{\bibinfo{volume}{98}},
  \bibinfo{pages}{084012} (\bibinfo{year}{2018}),
  \urlprefix\url{https://link.aps.org/doi/10.1103/PhysRevD.98.084012}.

\bibitem[{\citenamefont{Alcubierre and Torres}(2015)}]{Alcubierre:2014joa}
\bibinfo{author}{\bibfnamefont{M.}~\bibnamefont{Alcubierre}} \bibnamefont{and}
  \bibinfo{author}{\bibfnamefont{J.~M.} \bibnamefont{Torres}},
  \bibinfo{journal}{Class. Quant. Grav.} \textbf{\bibinfo{volume}{32}},
  \bibinfo{pages}{035006} (\bibinfo{year}{2015}), \eprint{1407.8529}.

\bibitem[{\citenamefont{Ruiz et~al.}(2012)\citenamefont{Ruiz, Degollado,
  Alcubierre, N\'u\~nez, and Salgado}}]{PhysRevD.86.104044}
\bibinfo{author}{\bibfnamefont{M.}~\bibnamefont{Ruiz}},
  \bibinfo{author}{\bibfnamefont{J.~C.} \bibnamefont{Degollado}},
  \bibinfo{author}{\bibfnamefont{M.}~\bibnamefont{Alcubierre}},
  \bibinfo{author}{\bibfnamefont{D.}~\bibnamefont{N\'u\~nez}},
  \bibnamefont{and} \bibinfo{author}{\bibfnamefont{M.}~\bibnamefont{Salgado}},
  \bibinfo{journal}{Phys. Rev. D} \textbf{\bibinfo{volume}{86}},
  \bibinfo{pages}{104044} (\bibinfo{year}{2012}),
  \urlprefix\url{https://link.aps.org/doi/10.1103/PhysRevD.86.104044}.

\bibitem[{\citenamefont{Jim\'enez-V\'azquez and
  Alcubierre}(2022)}]{PhysRevD.105.064071}
\bibinfo{author}{\bibfnamefont{E.}~\bibnamefont{Jim\'enez-V\'azquez}}
  \bibnamefont{and}
  \bibinfo{author}{\bibfnamefont{M.}~\bibnamefont{Alcubierre}},
  \bibinfo{journal}{Phys. Rev. D} \textbf{\bibinfo{volume}{105}},
  \bibinfo{pages}{064071} (\bibinfo{year}{2022}),
  \urlprefix\url{https://link.aps.org/doi/10.1103/PhysRevD.105.064071}.

\bibitem[{\citenamefont{Lai and Choptuik}(2007)}]{Lai:2007tj}
\bibinfo{author}{\bibfnamefont{C.~W.} \bibnamefont{Lai}} \bibnamefont{and}
  \bibinfo{author}{\bibfnamefont{M.~W.} \bibnamefont{Choptuik}}
  (\bibinfo{year}{2007}), \eprint{0709.0324}.

\bibitem[{\citenamefont{Bernal}(2021)}]{BernalA_2021}
\bibinfo{author}{\bibfnamefont{A.}~\bibnamefont{Bernal}},
  \bibinfo{howpublished}{private communication} (\bibinfo{year}{2021}).

\end{thebibliography}


\end{document}